\documentclass[prb,showpacs,twocolumn,groupedaddress,floatfix]{revtex4}

\usepackage{amsmath,amssymb}
\usepackage{graphics}

\begin{document}

\title{Canyon of Current Suppression in an interacting two-level Quantum Dot}
\author{O. Karlstr\"om, J. N. Pedersen\cite{Pedersenaffiliation}, P. Samuelsson, A. Wacker}

\affiliation{Mathematical Physics, University of Lund, Box 118, 22100 Lund, Sweden}

\date{\today,~accepted by PRB}

\begin{abstract}
Motivated by the recent discovery of a canyon of conductance suppression in a two-level equal spin quantum dot system 
[Phys. Rev. Lett. $\bf{104}$, 186804 (2010)] the transport through this system is studied in detail. 
At low bias and low temperature a strong current suppression is found around
the electron-hole symmetry point independent of the couplings, in agreement with
previous results. By means of a Schrieffer-Wolff transformation
we are able to give an intuitive explanation to this suppression in the low-energy regime.
In the general situation, numerical simulations are carried out using quantum rate equations. 
The simulations allow for the prediction of 
how the suppression is affected by the couplings, the charging energy, 
the position of the energy levels, the applied bias, and the temperature. 
We find that away from electron-hole symmetry, the parity of the couplings is essential
for the current suppression. It is also shown how broadening, interference,
and a finite interaction energy cause a shift of the current minimum away from degeneracy.
Finally we see how an increased population of the upper level leads to
current peaks on each side of the suppression line. At sufficiently high bias we discover a coherence-induced 
population inversion.
\end{abstract}


\maketitle

\section{INTRODUCTION}
The conductance of quantum dots is dominated by their discrete level spectra.\cite{FranceschiPRL2001,BeenakkerPRB1991} 
Particularly interesting features can 
be found at degeneracies of the discrete levels.\cite{OosterkampPRL1998,ReimannRMP2002,PayettePRB2010} 
At zero magnetic field spin-degenerate levels can be modeled with 
the Anderson Hamiltonian.\cite{AndersonPR1961} At low temperatures such systems exhibit the intriguing phenomenon of the Kondo effect,
which has been extensively investigated.
\cite{PustilnikJPC2004} On the contrary, much less is known about systems with orbital degeneracies. The simplest system with such 
degeneracies is the two-level spinless quantum dot, which was studied theoretically in connection with phenomena such as
phase lapses of the transmission phase,\cite{PhysRevB.55.13726,SilvaPRB2002,GolosovPRB2006} charge oscillations, 
\cite{PhysRevB.71.201308,SindelPRB2005} and correlation induced conduction resonances.\cite{MedenPRL2006} 
Relations between these phenomena were discussed in Refs.~\onlinecite{KashcheyevsPRB2007, HyunPRL2007}.
While Refs.~\onlinecite{PhysRevB.55.13726, SilvaPRB2002, GolosovPRB2006, PhysRevB.71.201308, SindelPRB2005, MedenPRL2006, KashcheyevsPRB2007,HyunPRL2007},  
investigated the properties of the system in the low-bias limit, the present paper deals with the effects of finite bias and temperature. 
For typical parameters applicable to experiments, we find that changing the bias might lead to qualitatively very different results. 
Finite bias was previously investigated in Ref.~\onlinecite{SchallerPRB2009}, where the current and full counting statistics 
for the system were studied, but only to lowest order in lead-dot coupling. The effect of finite bias and temperature in
the co-tunneling regime has thus not previously been investigated. 
A two-level quantum dot system, where the two levels
were coupled to separate source and drain contacts, were studied in a recent paper.\cite{TrochaPRB2010} Here the orbital
quantum number has the same effect as spin, due to the separate contacts, and orbital Kondo effect can be investigated. The paper studied
the level renormalization, Kondo temperature, local density of states and conductance of the system.

On the experimental side, the degeneracy of orbital levels was recently studied for a gate-defined quantum dot in an InSb 
nanowire,\cite{NilssonPRL2010} where a 
canyon of conductance suppression was found. Here large level-dependent g-factors enabled the study of degenerate
orbital levels using the Zeeman effect.\cite{NilssonNL2009} Similar systems of quantum dots embedded in nanowires have previously been 
experimentally realized.\cite{BjorkNano2004,ShorubalkoNT2006,HuNatureNanotech2007} The model is not restricted to
the two-level quantum dot but can also be applied to parallel quantum dots coupled to the same leads.\cite{PhysRevLett.87.256802,PhysRevLett.92.176801,PhysRevLett.93.066802}

\begin{figure}[ht]
\begin{center}
\resizebox{!}{40mm}{\includegraphics{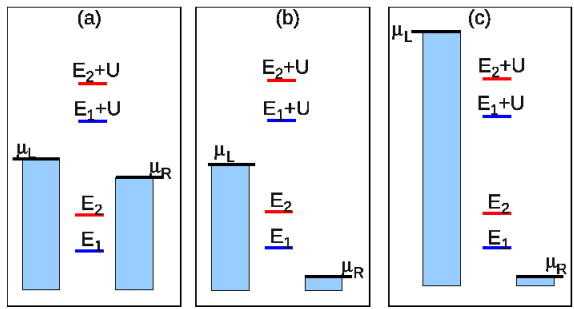}}
\end{center}
\caption{Level configuration at low bias (a), high bias (b) and ultra-high bias (c).
Due to the Coulomb interaction the energies of the levels are increased by an
amount $U$ when both states are filled. The applied bias is given by $\mu_L-\mu_R$,
where $\mu_L$ and $\mu_R$ are the chemical potentials in the left and right contact, respectively.}
\label{electronholsym}
\end{figure}

Previous theoretical investigations of the equal spin two-level quantum dot showed the existence of a complete conductance suppression 
around the electron-hole
symmetry point in the limit of zero temperature.\cite{MedenPRL2006, KashcheyevsPRB2007, SilvestrovPRB2007,HyunPRL2007} 
In this paper we investigate the current suppression at finite bias and temperature, as well as the current away from 
electron-hole symmetry, i.e. regimes not previously studied beyond sequential tunneling. 
The aim of this paper is to determine whether the canyon of 
current suppression is generic or appears 
only for certain parameters. In the regime $\Gamma >V_{\mathrm{bias}},~k_BT$, where $\Gamma$ is the level broadening given by 
the coupling between dot levels and leads, we find a complete current suppression close to degeneracy, both in the
co-tunneling and the sequential tunneling regime, assuming that the two dot states couples to the leads with different parity. For 
weaker couplings to the leads, or for increased $V_{\mathrm{bias}}$ or $k_BT$, only partial current suppression is found.
Results are compared with three limiting cases of bias. i)~The low-bias limit, $V_{\mathrm{bias}}<\Gamma,U$, is shown in 
Fig.~\ref{electronholsym}(a),
where $U$ is the charging energy.
ii)~The high-bias limit 
$\Gamma,\Delta E<V_{\mathrm{bias}}<U$, see Fig.~\ref{electronholsym}(b), where $\Delta E$ is the splitting between the two levels, 
so that both $E_1$ and $E_2$ are 
entirely inside the bias window. iii)~The ultra-high-bias in which the bias is the largest parameter 
$\Gamma,\Delta E,U<V_{\mathrm{bias}}$, as shown in Fig.~\ref{electronholsym}(c).

For the high- and ultra-high-bias limit the current can be evaluated by analytical means, as first-order approximations in tunneling 
are valid. We find that in the high-bias limit,
the current suppression is completely independent of the couplings.\cite{GurvitzEPL2009,PhysRevB.80.033302} 
In the ultra-high-bias limit, on the other hand, only partial suppression is found when the two states have different coupling strengths 
to the leads.

The remainder of the article is organized as follows.
The system Hamiltonian and the modeling of transport is discussed in Sec.~II.
The canyon of current suppression is studied in Sec.~III using the second order 
von Neumann method (2vN), where co-tunneling and coherence are included.\cite{PedersenPRB2005extra} 
The findings of this 
section are further explained in Sections~IV-VII. In Sec.~IV an intuitive explanation for the conductance suppression is 
given by means of a Schrieffer-Wolff transformation.\cite{SchriefferPR1966} In Sec.~V the current at electron-hole symmetry
is analyzed as a function of bias, temperature, and couplings. The current away from electron-hole symmetry is studied in Sec.~VI.
In Sec.~VII focus is changed to the current peaks surrounding the 
canyon of current suppression. At increased bias we discover an unexpected population inversion
and investigate its importance for the peaks. Here the term population inversion means a higher population of the
upper level. It should be pointed out that this population inversion results from a finite bias and is different from what is studied in
Refs.~\onlinecite{PhysRevB.71.201308,SindelPRB2005}.

\section{Modeling of transport}
In this section we discuss the Hamiltonian of the spinless two-level quantum dot as well as the modeling of the current 
through the system. In the single-particle eigenbasis of the dot, the system Hamiltonian is given by:

\begin{eqnarray}
\hat{H}&=&\hat{H}_{\mathrm{dot}}+\hat{H}_{\mathrm{leads}}+\hat{H}_{T},\\
\hat{H}_{\mathrm{dot}}&=&E_{1}d_{1}^{\dagger}d_{1}+E_{2}d_{2}^{\dagger}d_{2}+Ud_{1}^{\dagger}d_{1}d_{2}^{\dagger}d_{2}, \label{U} \\
\hat{H}_{\mathrm{leads}}&=&\sum_{k,\ell=L/R}E_{k}c_{k\ell}^{\dagger}c_{k\ell},\\
\hat{H}_{T}&=&\sum_{k,\ell=L/R}(t_{\ell1}d_{1}^{\dagger}+t_{\ell2}d_{2}^{\dagger})c_{k\ell}+\mathrm{H.c.},\label{HT}
\end{eqnarray}

\noindent where we have assumed that the couplings $t_{\ell i}$ are independent of $k$ and 
$\Gamma_{\ell i}(E)=2\pi t_{\ell i}^{2}\rho_0$ with a constant density of states $\rho_0(E)=\sum_k\delta(E_k-E)$, for $-D<E_k<D$. 
In the simulations a large bandwidth $D$ is used, assuming wide conduction bands of the leads.
The operators $d_i$ ($d_i^{\dagger}$) and $c_{kl}$ 
($c_{kl}^{\dagger}$) are annihilation (creation) operators of electrons in the dot and leads, respectively. 
In Eq.~(\ref{U}) the charging energy $U$ is due to Coulomb interaction between the electrons
when both dot states are filled. 

The energy levels are parametrized by

\begin{eqnarray}
E_{1/2}=\pm\frac{\Delta E}{2}-E_g-U/2. \label{levels}
\end{eqnarray}

\noindent Electron-hole symmetry, where the levels are symmetrically placed around the Fermi energy, is given by $E_g=0$ and $\Delta E=0$.
At this point electrons and holes contribute equally to the current through the quantum dot. 
Bias is applied symmetrically to the quantum dot so that the chemical potentials are $\mu_L=V_{\mathrm{bias}}/2$ and 
$\mu_R=-V_{\mathrm{bias}}/2$, for the left and right contact, respectively.

In this paper we mainly study couplings of the type

\begin{eqnarray}
\label{couplings}
t_{L1}=t, ~~t_{R1}=t, ~~t_{L2}=-at, ~~t_{R2}=at,
\end{eqnarray}

\noindent where the asymmetry parameter $a$ is chosen to be real. Time-reversal symmetry assures that the coupling elements
are real, implying that the two dot states couple to the leads with the same or opposite parity.\cite{SilvaPRB2002}
This paper focuses on the case of opposite parity as it is 
essential for the canyon of current suppression at finite bias, see Sec.~III where also the effects of asymmetric couplings to
left and right leads are briefly discussed. Furthermore, it was shown in Ref.~\onlinecite{NilssonPRL2010} that good agreement between 
theory and experiment could be achieved for real couplings. As the broadening of the levels is 
proportional to the square of the coupling elements we have 
$\Gamma_{L1}=\Gamma_{R1}=\Gamma_1,~\Gamma_{L2}=\Gamma_{R2}=a^{2}\Gamma_1$.

The current through the dot is calculated using the second order von Neumann (2vN) method.\cite{PedersenPRB2005extra} 
This method is an extension of the generalized master equation. Starting from the von Neumann equation an equation of motion is derived 
not only for the reduced density matrix but also for the elements of the total density matrix consisting of a single electron-hole 
excitation in the contact. 
In the latter system of equations two electron-hole excitation elements, corresponding to co-tunneling, enter. In the time evolution of the 
co-tunneling processes there are contributions from three electron-hole excitation elements but these are ignored. This results in a closed
set of equations, which can be solved for the occupations of the dot levels, the coherences between these resulting from the coupling
to the leads, and the current flowing through the quantum dot.

There are currently several other methods that account for tunneling of higher order than sequential processes. 
The most widely used technique is the generalized master equation approach which can be derived in many different ways including 
the real-time diagrammatic technique\cite{KonigPRL1997,KonigPRB1998} and the Bloch-Redfield approach originally developed in
Refs.~\onlinecite{PhysRev.89.728,PhysRev.105.1206,RedfieldAMR1965}. Comparisons 
of different approaches have been performed in Refs.~\onlinecite{TimmPRB2008,Koller2010}. The detailed relation between 
these methods and the 2vN-method will be the subject of future investigations, where we show that the 2vN-method is similar 
to other methods of $4^{th}$ order in the couplings $t$, but contains certain
diagrams up to infinite order. For the Anderson model with infinite $U$ it has been shown to give the same current
as the resonant-tunneling approximation\cite{KonigDiss1998} in the real-time diagrammatic approach.\cite{PedersenPRB2007} This 
method includes irreducible 
diagrams with an arbitrary number of correlated tunneling processes, but restricted to one electron-hole excitation at any
given moment.

\section{THE CANYON OF CONDUCTANCE SUPPRESSION}

\begin{figure}[h!]
\begin{center}
\resizebox{!}{175mm}{\includegraphics{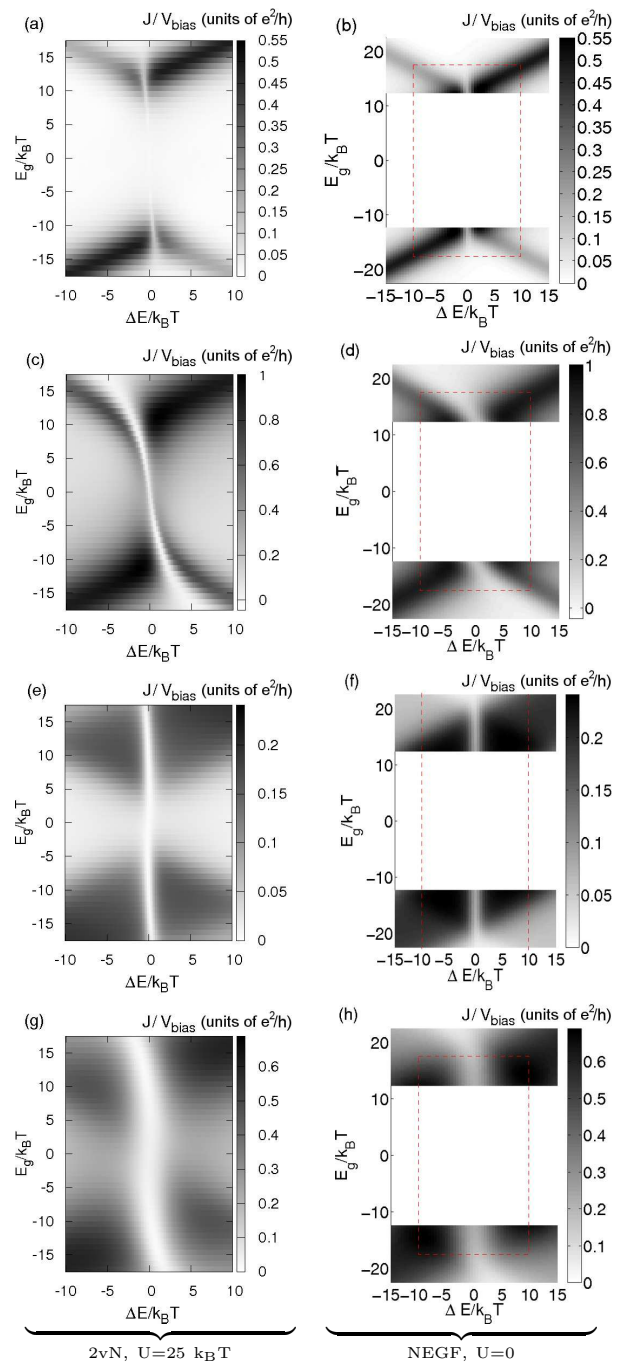}}
\begin{minipage}{.45\linewidth}
$\underbrace{~~~~~~~~~~~~~~~~~~~~~~~~~~~~~~~~~}_\mathrm{2vN,~U=25~k_BT}$
\newline
\end{minipage}
\begin{minipage}{.45\linewidth}
$\underbrace{~~~~~~~~~~~~~~~~~~~~~~~~~~~~~~~~~}_{\mathrm{NEGF,~U=0}}$
\newline
\end{minipage}
\caption{Conductance calculated with the 2vN method for the interacting system, $U=25~k_BT$ to the left, and noninteracting Green's 
function result to the
right. The white areas in the right figures are the regions where the noninteracting formalism is not valid. 
The upper four figures correspond to the low-bias limit $V_{\mathrm{bias}}=k_BT$, while the lower four show the results for high
bias $V_{\mathrm{bias}}=15~k_BT$. The couplings are given by Eq.~(\ref{couplings}) with $\Gamma_1=k_BT$ and $a=0.5$ for 
figures (a), (b), (e) and (f), while 
$\Gamma_1=4~k_BT$ and $a=0.5$ for figures (c), (d), (g) and (h). The dashed boxes in the noninteracting figures mark the studied region 
in the interacting case.} %CHANGED FROM $\Gamma=5$ to $\Gamma=4$.}
%k_BT=0.2
\label{canyon}
\end{center}
\end{figure}

We proceed by investigating whether the canyon of current suppression observed in Ref.~\onlinecite{NilssonPRL2010} can be found only 
in a certain parameter space or if it is
generic for the studied two-level system.
We plot the normalized current 
$J/V_{\mathrm{bias}}$, which equals the conductance $G$ in the low-bias limit.
As a start, we restrict to couplings given by Eq.~(\ref{couplings}) and study two different
coupling strengths $\Gamma_1/k_BT=1$ and $\Gamma_1/k_BT=4$. In both cases the relative coupling 
strength of the two levels is given by $a=0.5$. The reason for using $\Gamma_1>k_BT$ is that we want to study the current 
suppression in the co-tunneling regime. The complete suppression found close to degeneracy in the case of $\Gamma_1/k_BT=4$ is also 
supported by the analytical arguments of Sec. IV. Furthermore the 2vN-method is non-perturbative and contains some classes of diagrams
up to infinite order, which makes the analysis in the regime $\Gamma>k_BT$ more reliable.  
Furthermore, both low bias $V_{\mathrm{bias}}/k_BT=1$ and high bias 
$V_{\mathrm{bias}}/k_BT=15$ are studied to investigate the effects of finite bias. 
In the left part of Fig.~\ref{canyon}, the results of the 2vN simulations for
$U/k_BT=25$
are shown. In order to study the role of the Coulomb interaction,
corresponding results for $U=0$ are shown in the right part of Fig.~\ref{canyon}.
Here we
applied the method of nonequilibrium Green's functions (NEGF)\cite{SilvaPRB2002,RaikhPRB1994} providing identical
results to the numerically much more involved
2vN method (for $U=0$), for the simple system studied here.
We believe that this holds also for more complex 

\noindent systems, see also Refs.~\onlinecite{PedersenPRB2005extra, JinJCP2008}.

In order to facilitate the comparison,
we performed a cut of the noninteracting results at $E_g=0$ and
shifted the upper/lower part by $\pm 12.5 k_BT$. This mimics the presence
of Coulomb repulsion by one entirely filled or completely empty level, as
expected in the respective energy range. 
Plotting the noninteracting results
in this way we find indeed rather good agreement with the interacting results in the low bias regime,
allowing us to explain the essential features in this region. For high bias the current suppression is more pronounced in the 
interacting case.

The Coulomb blockade regime present in the left figures is a result of the interaction-related correlations. This regime can 
naturally not be studied using non-interacting methods.

The noninteracting results for weak bias, Fig.~\ref{canyon}(b,d), show  a line of
current suppression, which
is bowed towards the weaker resonance. 
This can be understood in terms of interference between the two levels. The Breit-Wigner formula provides

\begin{eqnarray}
T(E)=\Gamma_1^{2}\left|\frac{1}{E-E_1+i\Gamma_1}-\frac{a^{2}}{E-E_2+ia^{2}\Gamma_1}\right|^{2},
\label{Breit-Wigner}
\end{eqnarray}

\noindent and the current can then be calculated using

\begin{eqnarray}
J=\frac{1}{h}\int_{-\infty}^{\infty}[f_L(E)-f_R(E)]T(E)dE.
\end{eqnarray}

\noindent Vanishing conductance in the low temperature limit corresponds to $T(0)=0$, 
i.e. complete conductance suppression can only occur for $E_2=a^{2}E_1$. 
This explains the location of the strong current suppression found in Fig.~\ref{canyon}(b) and (d). 
It is easily shown that if one level couples symmetrically to left and right, complete suppression is only found
for couplings like Eq.~(\ref{couplings}), where the sign of $t_{Li}$ and $t_{Ri}$ differs for $i=1,2$. 
This motivates the special attention given to these couplings.

As the bias is increased the current not only depends on $T(0)$. When the bias is larger than the width of the dip, $a\Gamma_1$ at 
$E_1=E_2=0$ resulting from Eq.~(\ref{Breit-Wigner}), this results in only partial current suppression as can be seen 
in Fig.~\ref{canyon}(f, h).

Fig.~\ref{canyon}(a) shows that the weak coupling and low-bias interacting result is well reproduced by the noninteracting result, 
Fig.~\ref{canyon}(b), in 
its region of applicability ($\mid E_g \mid>U/2$), apart from a slight shift of the suppression line. In the Coulomb blockade regime 
($\mid E_g \mid<U/2$) almost no structure is
visible due to the very weak co-tunneling. In Sec.~V we see that this corresponds to the regime where $k_BT$ dominates and only partial
conductance suppression can be found.

Fig.~\ref{canyon}(c) shows that the 2vN-method can sometimes give unphysical results corresponding to small negative currents. 
This is expected as the 2vN method neglects some correlated
transitions of more than two electrons/holes. The contributions of these terms become important in the limit of large couplings
and small temperature and bias. Apart from this, the 2vN method agrees well with the noninteracting results. However, we see that the 
shift of the suppression line is larger when the couplings are strong. As explained in Sec.~VI this can be attributed to the broadening of the levels. 
In the Coulomb blockade regime a clear conductance suppression is found. This is the regime where $\Gamma$ dominates, 
i.e., the situation described by Refs.~\onlinecite{MedenPRL2006,KashcheyevsPRB2007,HyunPRL2007}.

In the high-bias limit, the interacting and noninteracting results differ significantly at the canyon of current suppression, 
as seen in Fig.~\ref{canyon}(e,~g) and Fig.~\ref{canyon}(f,~h), respectively. 
In the sequential tunneling  
regime a strong current suppression is found in the interacting case. This result has been explained by 
Refs.~\onlinecite{GurvitzEPL2009,PhysRevB.80.033302} and is
further discussed in Sec.~VI. This interaction-induced suppression is not included in the NEGF-method, resulting in only partial 
suppression, as can be seen from the gray colored canyon. 

The current suppression in the co-tunneling region becomes weaker as we are approaching the 
region where $V_{\mathrm{bias}}$ dominates, which 
results in only partial current suppression, see Sec.~V. This shows that the canyon of current suppression is found in all four cases
albeit the current not always entirely drops to zero. Further investigations show that the canyon of suppression is not restricted 
to couplings like Eq.~(\ref{couplings}), but is present also
for couplings which are asymmetric with respect to the left and right contact. 
The importance of the relative coupling strength, $a$, is investigated in Sec.~V.

\begin{figure}[ht]
\begin{center}
{\resizebox{!}{50mm}{\includegraphics{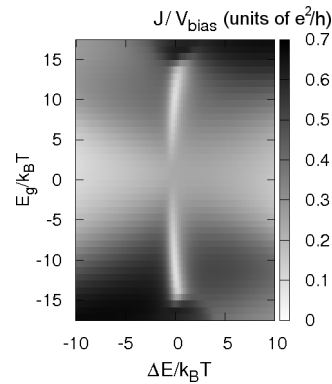}}}
\end{center}
\caption{Conductance calculated with the 2vN method. The two levels have equal parity with couplings given by 
$t_{L1}=t_{R1}=t$, $t_{L2}=0.4t$ and $t_{R2}=0.6t$ with $\Gamma_{L1}=\Gamma_{R1}=2\pi\rho_0t^2=5k_BT$.
Additional parameters are $V_{\mathrm{bias}}=15~k_BT$, $U=25~k_BT$.}
%k_BT=0.2
\label{samepar}
\end{figure}

For couplings where the two levels have the same parity, i.e. the two levels couple with the same sign to left and right lead, 
the situation is different, as shown in Fig.~\ref{samepar}. Here we still see the 
high-bias blockade in the sequential tunneling regime.\cite{GurvitzEPL2009,PhysRevB.80.033302} 
For $E_g<-U/2$ or $E_g>U/2$ interactions are of less 
importance and the blockade disappears. One also sees that no canyon is visible in the Coulomb blockade regime. The reason
is that the width of the dip in the transmission function becomes increasingly narrow as the singular coupling point is approached. 
At such couplings, one can at degeneracy completely decouple one level from the leads, resulting in a bistable system, see also 
Refs.~\onlinecite{MedenPRL2006,KashcheyevsPRB2007}. As a
result, the canyon is present only for very low bias and temperature, i.e. a regime where the 2vN-method is not applicable. 
Fig.~\ref{samepar} clearly shows that at finite bias the parity of the couplings is essential for the suppression, 
unlike the zero-bias limit.
Two levels with the same parity will not be studied any further. The remainder of the paper is devoted to the couplings described by 
Eq.~(\ref{couplings}). For such couplings, details of the canyon of current suppression is summarized in Table~I, which can be
of use for futher experimental studies.

\begin{table}[h]
\caption{A summary of the current suppression from Fig.~\ref{canyon}. For levels with couplings of different parity the current suppression is generic, 
but not always complete. The high-bias limit of Fig.~\ref{electronholsym}(b) implies to $E_g\approx \pm U/2$. Therefore the co-tunneling regime is not applicable 
(n.a.) in this limit. In the ultra-high-bias limit the current is given by the sequential tunneling result. The co-tunneling regime
is thus not categorized in this limit. The table also indicates in which sections the different regimes are investigated further.
}
\begin{center}
\begin{tabular}{| l | c | c | r | }
  \hline                       
  $~~~\Gamma$ & $V_{\mathrm{bias}}$ & co-tunneling & sequential tunneling \\
  &  & ($E_g\approx 0$) & ($E_g\approx \pm U/2$)~~~~ \\ \hline
  $>k_BT$ & low & complete & complete (Sec.~VI)\\
   &  & (Sec.~IV and V)& \\ \hline
  $>k_BT$ & high & n.a. & complete (Sec.~VI)\\ \hline
  $>k_BT$ & ultra-high & - & partial (Sec.~V)\\ \hline
  $<k_BT$ & low & partial (Sec.~V)& partial (Sec.~VI)\\ \hline
  $<k_BT$ & high & n.a. & complete (Sec.~VI)\\ \hline
  $<k_BT$ & ultra-high & - & partial (Sec.~V)\\
  \hline  
\end{tabular}
\end{center}
\end{table}

\section{TUNNELING PROCESSES IN THE LOW-ENERGY LIMIT}
In this section we study the current in the regime where $E_1,~E_2<0$ and $E_1+U,~E_2+U>0$, so that the dot is singly occupied. Furthermore we will
assume that $V_{\mathrm{bias}},~k_BT\rightarrow 0$. This allows us to derive an effective low-energy Hamiltonian by means of a 
Schrieffer-Wolff transformation. Low-energy implies that all tunneling events occur 
around $E_k=0$. The transformation requires that the charging 
energy of the dot is much larger than the broadening of the dot levels, so that the dot is singly occupied when the levels are 
placed as in Fig.~\ref{electronholsym}(a). By analyzing the effective Hamiltonian in this regime we will be able to identify the two-particle processes 
contributing to the 
current and give an intuitive explanation of the conductance suppression. It is convenient to write the Hamiltonian 
in terms of pseudo-spin operators\cite{KashcheyevsPRB2007,HyunPRL2007} defined by
$d_{\uparrow}=d_1$, $d_{\downarrow}=d_2$, $c_{k\uparrow}=(c_{kL}+c_{kR})/\sqrt{2}$ and
$c_{k\downarrow}=(-c_{kL}+c_{kR})/\sqrt{2}$.
The system Hamiltonian then reads

\begin{eqnarray}
\hat{H}_{\mathrm{dot}}&=&E_{\uparrow}d_{\uparrow}^{\dagger}d_{\uparrow}+E_{\downarrow}d_{\downarrow}^{\dagger}d_{\downarrow}+Ud_{\uparrow}^{\dagger}d_{\uparrow}d_{\downarrow}^{\dagger}d_{\downarrow},\\
\hat{H}_{\mathrm{leads}}&=&\sum_{k,\sigma=\uparrow/\downarrow}E_{k}c_{k\sigma}^{\dagger}c_{k\sigma},\\
\hat{H}_{T}&=&\sum_{\sigma k}t_\sigma c_{k\sigma}^{\dagger}d_\sigma+\mathrm{H.c.}~~~~\nonumber \\
&=&\sum_{k}\sqrt{2}t(c_{k\uparrow}^{\dagger}d_\uparrow+ac_{k\downarrow}^{\dagger}d_\downarrow)+\mathrm{H.c.}
\end{eqnarray}

\noindent Here $E_{\uparrow}=E_1$, $E_{\downarrow}=E_2$, $t_{\uparrow}=\sqrt{2}t$, $t_{\downarrow}=\sqrt{2}at$, 
and the problem has been mapped onto the one-lead Anderson model
with spin-dependent couplings. A similar transformation can be found for arbitrary couplings.

We now perform a Schrieffer-Wolff transformation, i.e. a canonical transformation

\begin{eqnarray}
\hat{H}_{S}=e^{iS}\hat{H}e^{-iS},
\end{eqnarray}

\noindent so that $\hat{H}_{S}$ does not have a linear term in $\hat{H}_{T}$. One finds that the following 
transformation $S=S^{+}+S^{-}$, with

\begin{eqnarray}
S^{-}=-i\sum_{\sigma k}\left(\frac{t_\sigma}{E_k-(E_\sigma+U)}n_{\overline{\sigma}}c_{k\sigma}^{\dagger}d_{\sigma}\right. \nonumber \\ 
\left.+\frac{t_\sigma}{E_k-E_\sigma}(1-n_{\overline{\sigma}})c_{k\sigma}^{\dagger}d_\sigma \right),
\end{eqnarray}

\noindent and $S^{+}=(S^{-})^{\dagger}$, will do the job. Here $\overline{\sigma}=\downarrow$ if $\sigma=\uparrow$ and $\overline{\sigma}=\uparrow$ if $\sigma=\downarrow$. 
Writing the tunneling Hamiltonian as $\hat{H}_{T}=\hat{H}_{T}^{+}+\hat{H}_{T}^{-}$, with

\begin{eqnarray}
\hat{H}_{T}^{-}=\sum_{\sigma k}t_\sigma c_{k\sigma}^{\dagger}d_\sigma,
\end{eqnarray}

\noindent and $\hat{H}_{T}^{+}=(\hat{H}_{T}^{-})^{\dagger}$, the effective low-energy Hamiltonian of the singly occupied subspace, when restricting to two-particle tunneling, is given by

\begin{eqnarray}
\hat{H}_{S}=&&\hat{H}_{\mathrm{dot}}+\hat{H}_{\mathrm{leads}}\nonumber \\
&&+\underbrace{\frac{i}{2}([S^{-},\hat{H}_{T}^{+}]+[S^{+},\hat{H}_{T}^{-}])}_{\hat{H}_{TS}}+O(\hat{H}_T^{3})
\end{eqnarray}

\noindent where the last term is the effective tunneling Hamiltonian $\hat{H}_{TS}$.
Performing the commutations and returning to our original basis one finds:

\begin{widetext}
\begin{eqnarray}
\label{lowenergy2}
\hat{H}_{TS}=&&\frac{1}{2}\overline{\sum_{kk'}}\left(\frac{t^{2}d_2^{\dagger}d_2}{E_k-(E_1+U)}+\frac{t^{2}(1-d_2^{\dagger}d_2)}{E_k-E_1}\right)(c_{kL}^{\dagger}+c_{kR}^{\dagger})(c_{k'L}+c_{k'R})~~~~~~ \nonumber \\
&+&\frac{1}{2}\overline{\sum_{kk'}}\left(\frac{a^{2}t^{2}d_1^{\dagger}d_1}{E_k-(E_2+U)}+\frac{a^{2}t^{2}(1-d_1^{\dagger}d_1)}{E_k-E_2}\right)(c_{kL}^{\dagger}-c_{kR}^{\dagger})(c_{k'L}-c_{k'R}) ~\\
&+&\frac{1}{2}\sum_{kk'}\left(\frac{at^{2}}{E_k-(E_1+U)}-\frac{at^{2}}{E_k-E_1}\right)(c_{kL}^{\dagger}+c_{kR}^{\dagger})(c_{k'L}-c_{k'R})d_2^{\dagger}d_1~~~ \nonumber \\
&+&\frac{1}{2}\sum_{kk'}\left(\frac{at^{2}}{E_k-(E_2+U)}-\frac{at^{2}}{E_k-E_2}\right)(c_{kL}^{\dagger}-c_{kR}^{\dagger})(c_{k'L}+c_{k'R})d_1^{\dagger}d_2~~~ \nonumber \\
&+&\mathrm{H.c.}+\hat{H}_{\mathrm{renorm}}~~~~~~~~~~~~~~~~~~~~~~~~~~~~~~~~~~~~~~~~~~~~~~~~~~~~~~~~~~~~~~~~~~~~~~ \nonumber 
\end{eqnarray}
\end{widetext}

\noindent where $\overline{\sum}$ indicates that the terms containing $c_{kL}^{\dagger}c_{kL}+c_{kR}^{\dagger}c_{kR}$, 
corresponding to the renormalization of energy levels $H_{\mathrm{renorm}}$, are not included in the sum.
In the limit of zero temperature and zero bias, the renormalization reads (see Appendix B)

\begin{eqnarray}
\label{renormalization}
\hat{H}_{\mathrm{renorm}}=\frac{\Gamma_1}{\pi}\left(\mathrm{ln}\left| \frac{E_1}{E_1+U} \right|d_1^{\dagger}d_1\right. \nonumber \\
\left.+a^{2}\mathrm{ln}\left| \frac{E_2}{E_2+U} \right|d_2^{\dagger}d_2\right).
\end{eqnarray}

\noindent It should be noted that the above expression 
does not hold if one of the levels is close to the chemical potential, as the Schrieffer-Wolff transform breaks down in this regime. 
The effect of the renormalization will be further discussed in Sec.~V, here we simply note that at electron-hole symmetry, 
where $E_1=E_2=-U/2$, the renormalization vanishes. 

Each term in Eq.~(\ref{lowenergy2}) corresponds 
to a co-tunneling process, 
where the electron is transferred from one lead to the other, or returns to the original lead. 
The first two sums correspond to elastic processes 
where the state of the dot is unchanged. The next two sums correspond to inelastic processes where the state of the dot changes.

Generally there is an effective overlap matrix element between the two dot states originating from the coupling to the leads. 
In Refs.~\onlinecite{BoesePRB2001,SilvestrovPRB2007} it was shown that for couplings of the type Eq.~(\ref{couplings}), 
this overlap vanishes. Indeed, evaluating
the $k=k'$ inelastic terms in Eq.~(\ref{lowenergy2}) results in factors $c_{kL}^{\dagger}c_{kL}-c_{kR}^{\dagger}c_{kR}=0$ 
in the low-bias limit, see Sec. VII for a discussion concerning finite bias.

Let us now analyze Eq.~(\ref{lowenergy2}) in detail: We see that only the inelastic processes affect the occupations of the dot, as $d_1^{\dagger}d_1$ 
and $d_2^{\dagger}d_2$ commutes with the other parts of $\hat{H}_{TS}$.
For $E_1=E_2$ the third and fourth sum in Eq.~(\ref{lowenergy2}), corresponding to the inelastic processes, are symmetric in level 1 and 2 except 
for a sign change $\mp c_{kL}^{\dagger}c_{k'R}\pm c_{kR}^{\dagger}c_{k'L}$.
In the low-bias limit, this sign change does not affect the kinetics of the occupations and we conclude that both levels are half filled at degeneracy.

The current operator reads $\hat{J}_{S}=-\sum_k \frac{i}{\hbar}[\hat{H_S},c_k^{\dagger}c_k]$, resulting in

\begin{widetext}
\begin{eqnarray}
\label{currentoperator}
\hat{J}_{S}=&-&\frac{it^{2}}{2\hbar}\sum_{kk'}\left(\frac{d_2^{\dagger}d_2}{E_k-(E_1+U)}+\frac{d_1^{\dagger}d_1}{E_k-E_1}-\frac{a^{2}d_1^{\dagger}d_1}{E_k-(E_2+U)}-\frac{a^{2}d_2^{\dagger}d_2}{E_k-E_2} \right)(c_{kR}^{\dagger}c_{k'L}-c_{kL}^{\dagger}c_{k'R}) \\
&-&\frac{iat^{2}}{\hbar}\sum_{kk'}\left[\left(\frac{1}{E_k-(E_1+U)}-\frac{1}{E_k-E_1} \right)d_2^{\dagger}d_1-\left(\frac{1}{E_k-(E_2+U)}-\frac{1}{E_k-E_2} \right)d_1^{\dagger}d_2\right] (c_{kR}^{\dagger}c_{k'L}+c_{kL}^{\dagger}c_{k'R}) \nonumber \\
&+&\mathrm{H.c.}, \nonumber
\end{eqnarray}
\end{widetext}

\noindent where $d_1^{\dagger}d_1+d_2^{\dagger}d_2=1$ has again been used. The first row corresponds to elastic processes, while the
second row constitutes the inelastic processes. We adopt the notation $\left.\vline~0\right\rangle$, $\left.\vline~1\right\rangle$, 
$\left.\vline~2\right\rangle$, and $\left.\vline~d\right\rangle$, corresponding to the dot being empty, in state 1, in state 2, or doubly 
occupied. The term
$d_2^{\dagger}d_2/(E_k-(E_1+U))$ represents
the elastic processes $\left.\vline~2\right\rangle \rightarrow \left.\vline~d\right\rangle \rightarrow \left.\vline~2\right\rangle$. 
The denominator includes the energy difference between the two states 
involved in the process, $E_d-E_2=E_1+E_2+U-E_2=E_1+U$, and the numerator represents the fact that state $\left.\vline~2\right\rangle$ 
has to be occupied 
for this process to occur. In the same way the term $d_1^{\dagger}d_1/(E_k-E_1)$ represents the process 
$\left.\vline~1\right\rangle \rightarrow \left.\vline~0\right\rangle \rightarrow \left.\vline~1\right\rangle$.
Assuming $E_k=0$, corresponding to low bias and low temperature, we see that these
terms cancel completely at the electron-hole symmetry point $E_1=E_2=-U/2$, owing to the equal population of the two levels. We note 
that these two processes, involving tunneling into and out of state $\left.\vline~1\right\rangle$, use the couplings of this state. 
In the same way we
see how the two processes $\left.\vline~1\right\rangle \rightarrow \left.\vline~d\right\rangle \rightarrow \left.\vline~1\right\rangle$ 
and $\left.\vline~2\right\rangle \rightarrow \left.\vline~0\right\rangle \rightarrow \left.\vline~2\right\rangle$ cancel.
The factor of $a^{2}$ enters here as it is state $\left.\vline~2\right\rangle$ that is active in these processes. We have seen that the elastic co-tunneling
processes cancel completely at the electron-hole symmetry point, independent of the relative coupling strength of the two levels.
Away from electron-hole symmetry, we see that the processes no longer cancel. However, at $E_1=E_2$, we again find complete 
canceling for $a=1$.

We now investigate the inelastic processes. It can easily be seen that the terms involving $d_2^{\dagger}d_1$ and $d_1^{\dagger}d_2$ cancel
at $E_1=E_2$, due to the equal occupation of the two levels. Electron-hole symmetry is thus not required for the canceling of the
inelastic terms, degeneracy of the two levels is sufficient. 

To summarize we have shown that at $E_1=E_2=-U/2$ the current suppression is complete, in the limit of low bias 
and low temperature at the electron-hole symmetric point. Unlike previous explanations of this phenomena our derivation gives an 
intuitive explanation to the suppression in 
terms of canceling of co-tunneling processes. It also explains why the suppression is not complete for finite bias and temperature or away
from electron-hole symmetry. This will be further studied in Secs.~V and VI respectively. 
The suppression is not an interference-effect between different processes, since the processes that cancel involve 
different initial and final states. The processes instead cancel as a result of the equal population of the two levels. 
At degeneracy, but away from electron-hole symmetry, the elastic processes do not cancel unless $a=1$. There is, however, 
partial canceling due to the different parity of the two levels. We stress that the complete suppression at electron-hole symmetry is
not an effect of this difference in parity. The pseudo-spin operators are always orthogonal. It is this 
orthogonality that gives equal occupations of the two levels and ensures the canceling of current carrying processes in 
the low-bias limit. The canceling is due to the phase lapse between the elastic co-tunneling processes using $\left.\vline~0\right\rangle$ and 
$\left.\vline~d\right\rangle$ as the virtual intermediate state. This shows the relationship between population switches, phase lapses,
and vanishing conductance, which was previously investigated in Ref.~\onlinecite{KashcheyevsPRB2007}. At electron-hole symmetry
the vanishing conductance is thus an effect of the correlations induced by the Coulomb interaction.

\section{CURRENT SUPPRESSION AT ELECTRON-HOLE SYMMETRY}

In the previous section we saw that the current suppression was most pronounced at the electron-hole symmetry point. 
In this section we will 
investigate the suppression at this point in more detail. The current is calculated as the flow of particles from the left contact
into the dot. Due to current conservation this is, in the stationary case, of course equal to the current flowing from the dot into 
the right contact. Thus, we do not label the current with any lead-index. To quantify the extent of the current suppression we introduce 
the dimensionless ratio between the current through
the two-level dot corresponding to Eq.~(\ref{U}) and the current through a single level dot containing only level one

\begin{eqnarray}
Q=\frac{J_{\mathrm{both~levels}}}{J_{\mathrm{level~1}}}.
\end{eqnarray}

\noindent This ratio measures how much the presence of level 2 suppresses the current. 
In the regime where sequential tunneling is dominating
the first order von Neumann  method (1vN) can be used to approximately evaluate the current, see Appendix A.
Using the couplings of Eq.~(\ref{couplings}), it is possible to derive Eq.~(\ref{current2}). At electron-hole symmetry, $E_1=E_2=-U/2$, 
this simplifies to

\begin{widetext}
\begin{eqnarray}
J=&&\frac{\Gamma_1}{\hbar}\left\lbrace w_{00}f_L(-U/2)(1+a^2)-w_{11}[1-f_L(-U/2)]\right.
\left. +w_{11}f_L(U/2)a^2-w_{22}[1-f_L(-U/2)]a^2 \right. \nonumber \\
&&\left.+w_{22}f_L(U/2)-w_{dd}[1-f_L(U/2)](1+a^2)+2a\mathcal{R}\left\lbrace w_{12}\right\rbrace [1-f_L(-U/2)+f_L(U/2)] \right\rbrace,
\label{Peter}
\end{eqnarray}
\end{widetext}

\noindent where $w_{00}$, $w_{11}$, $w_{22}$ and $w_{dd}$ is the probability to find the dot empty, in state 1, in state 2 and 
doubly occupied respectively, 
while $w_{12}$ is the coherence between level 1 and 2. Here $\mathcal{R}$ denotes the real part. In this paper we refer to the coherence as the non-diagonal elements of the 
reduced density matrix. The coherence is thus a basis-dependent quantity.
The first terms of Eq.~(\ref{Peter}) are identical to the corresponding Pauli master equation and are easily interpreted in terms of transition rates. 
The last term originating from the coherence of the two levels is more difficult to understand intuitively. 
However, it is evident that a negative real part of the coherence results in a decreased current.

At level degeneracy one can get rid of the coherence term by performing a basis change that diagonalizes the reduced density matrix. 
Here we focus at the case of electron-hole symmetry, $E_1=E_2=-U/2$, where level 1 and 2 have the same occupation and the 
coherence is real (see Appendix A). The singly occupied part of the reduced density matrix can thus be written

\[ \left( \begin{array}{cc}
w_{11} & w_{12}\\
w_{21} & w_{22}\\
\end{array} \right) =\left( \begin{array}{cc}
w_{11} & w_{12}\\
w_{12} & w_{11}\\
\end{array} \right)=SDS^{-1}. \]

\noindent The diagonalization is given by

\[ S=\frac{1}{\sqrt{2}}\left( \begin{array}{cc}
1 & 1\\
1 & -1\\
\end{array} \right),~D=\left( \begin{array}{cc}
w_{11}+w_{12} & 0\\
0 & w_{11}-w_{12}\\
\end{array} \right)\]

\noindent where the diagonal elements of $D$ are the occupations in the new basis.

The couplings in the new eigenbasis are given by

\begin{eqnarray}
t_{L1}'=\frac{t}{\sqrt{2}}(1-a),~
t_{R1}'=\frac{t}{\sqrt{2}}(1+a),~ \\
t_{L2}'=\frac{t}{\sqrt{2}}(1+a),~
t_{R2}'=\frac{t}{\sqrt{2}}(1-a).\nonumber~
\label{newcouplings}
\end{eqnarray}

\noindent In the new basis one level couples strongly to 
the left, while the other couples strongly to the right lead. This results in a strong current suppression and also explains the 
occupations in our new basis: the level which is strongly
coupled to the lead with a high chemical potential has a high occupation. A high bias thus results in a large coherence between the
levels, cf. the expression for the matrix $D$. For $a=1$, i.e. both levels have the same coupling strength, 
each level decouples completely from one lead. As a result no current can flow through the quantum dot, see also Ref.~\onlinecite{BoesePRB2001}. 
The effect of the coherence is thus to decrease the current. As the coherence is zero in the new basis one can
use the Pauli master equation approach to derive the current. At electron-hole symmetry we obtain

\begin{eqnarray}
J_{\mathrm{both~levels}}=\frac{(1-a^{2})^{2}\Gamma_1}{2\hbar(1+a^{2})}[f_L(-U/2)-f_R(-U/2)].
\label{easycurrent}
\end{eqnarray}

\noindent If only level 1 was present the current would be given by

\begin{eqnarray}
J_{\mathrm{level~1}}=\frac{\Gamma_1}{2\hbar}[f_L(-U/2)-f_R(-U/2)],
\end{eqnarray}

\noindent and the resulting ratio is

\begin{eqnarray}
Q=\frac{J_{\mathrm{both~levels}}}{J_{\mathrm{level~1}}}=\frac{(1-a^{2})^{2}}{1+a^{2}}.
\label{highbias}
\end{eqnarray}

\noindent This result is valid when sequential tunneling is dominating. 

\begin{figure}[ht]
\begin{center}
\resizebox{!}{60mm}{\includegraphics{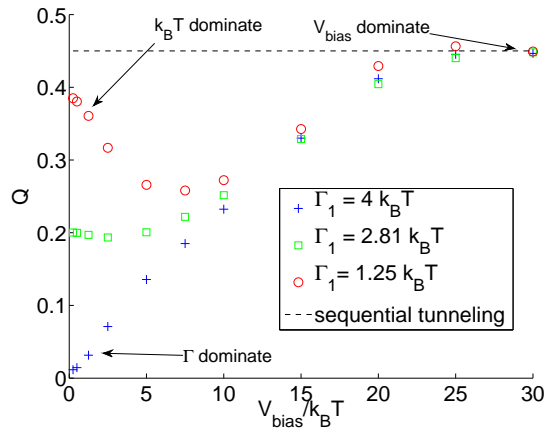}}
\end{center}
\caption{$Q$ as a function of $V_{\mathrm{bias}}$ at electron-hole symmetry for different coupling strengths. 
All curves correspond to $a=0.5$ and $U=25~k_BT$.}%$k_BT=0.2$
\label{QGamma}
\end{figure}

In Fig.~\ref{QGamma} the current 
suppression, $Q$, at electron-hole symmetry, $E_1=E_2=-U/2$, is shown as a function of the applied bias.
The numerator of $Q$ is calculated using the 2vN method, while the denominator is calculated using NEGF which, however, agrees exactly with
2vN for the single level dot.
Fig.~\ref{QGamma} clearly shows that when $V_{\mathrm{bias}}>U$, the ultra-high-bias regime, where sequential tunneling is dominating, 
is entered and the results of the 2vN simulations 
agree very well with the analytical results, Eq.~(\ref{highbias}). 

In the low-bias limit the results are more complicated. For weak couplings the contribution from co-tunneling processes is small
and the first-order result Eq.~(\ref{highbias}) is approached. 
As the bias is increased the phase space for co-tunneling processes ($\sim V_{\mathrm{bias}}$) increases, so these can
no longer be neglected. This results in a stronger current 
suppression and a related decrease in $Q$ for the weaker couplings. For strong couplings, i.e. $\Gamma>V_{\mathrm{bias}}, k_BT$, 
co-tunneling dominate and the low-energy Hamiltonian of Sec.~IV predicts complete suppression in the low-bias limit.  
We stress that the suppression in the low-bias limit is not an artifact of neglecting higher order processes. As was shown in Sec.~IV and
Refs.~\onlinecite{MedenPRL2006,KashcheyevsPRB2007,HyunPRL2007} and \onlinecite{SilvestrovPRB2007},
the suppression is complete in the limit of low bias and low temperature, at the electron-hole symmetric point.

\begin{figure}[t]
\begin{center}
\resizebox{!}{60mm}{\includegraphics{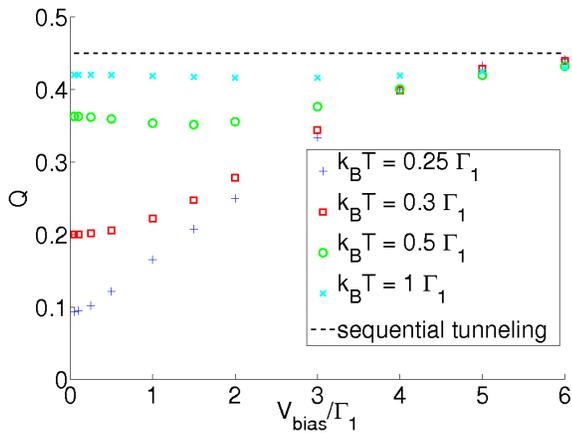}}
\end{center}
\caption{$Q$ as a function of $V_{\mathrm{bias}}$ at electron-hole symmetry for different temperatures. All curves correspond to 
$a=0.5$ and $U=5~\Gamma_1$.}%$\Gamma_1=1$
\label{QkBT}
\end{figure}

In Fig.~\ref{QkBT} the simulated current suppression at electron-hole symmetry is shown for different temperatures as a function of bias. As 
the temperature is not constant in Figs.~5 and 6 we here express all energies in terms of $\Gamma_1$. 
It is clearly seen that the sequential tunneling result is reached in the high-temperature limit.
%\vspace{20cm}

\begin{figure}[t]
\begin{center}
\resizebox{!}{60mm}{\includegraphics{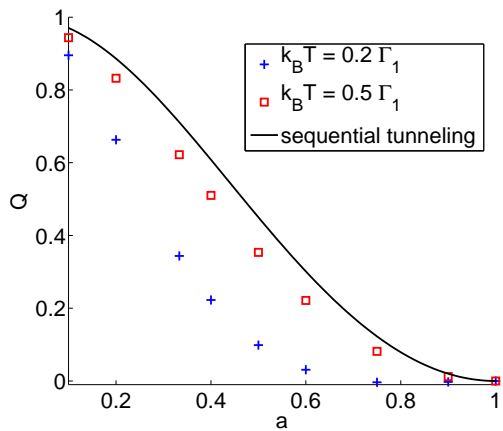}}
\end{center}
\caption{$Q$ as a function of the difference in coupling strength, $a$ see Eq.~(\ref{couplings}), 
at electron-hole symmetry for different temperatures. The high temperature limit is given by Eq.~(\ref{highbias}). 
All curves correspond to $V_{\mathrm{bias}}=1~\Gamma_1$ and $U=5~\Gamma_1$.}%$\Gamma_1=1$
\label{change_a}
\end{figure}

In Fig.~\ref{change_a} we investigate the current suppression $Q$ as a function of the difference $a$ in coupling strength between 
the two levels for different temperatures. For lower temperatures it can be seen that there is essentially complete suppression 
when both levels have equal coupling strength, i.e. $a\approx1$. When $a\approx0$ and the difference in coupling strength is large, 
almost no suppression is observed. This is expected as no current flows through the weakly coupled state, and due to electron-hole 
symmetry it does not matter if it is occupied or not.

In this Section we investigated how a finite bias and temperature affects the suppression at electron-hole symmetry.
In the limit of low/high bias and temperature we compared with previously established results/first order simulations, and found 
good agreement. In intermediate regimes we found that the current suppression could show a non-monotonous behavior, see 
$\Gamma_1=1.25~k_BT$ in Fig.~4. Furthermore Figs.~\ref{QGamma},~\ref{QkBT} and \ref{change_a} show that $Q$ increases 
with increasing temperature until the high temperature limit is reached. We have seen how the current suppression is given 
by analytical first order results in the 
regimes where sequential tunneling is dominating. In this regime the suppression is coherence-induced and only partial unless both 
levels have the same coupling strength. Stronger couplings lead to an increased co-tunneling which results in 
a stronger suppression.

\section{CURRENT SUPPRESSION AWAY FROM ELECTRON-HOLE SYMMETRY}

\begin{figure}[t!]
\begin{center}
\resizebox{!}{60mm}{\includegraphics{Fig7a.eps}}
\begin{minipage}{.45\linewidth}
\resizebox{!}{26mm}{\includegraphics{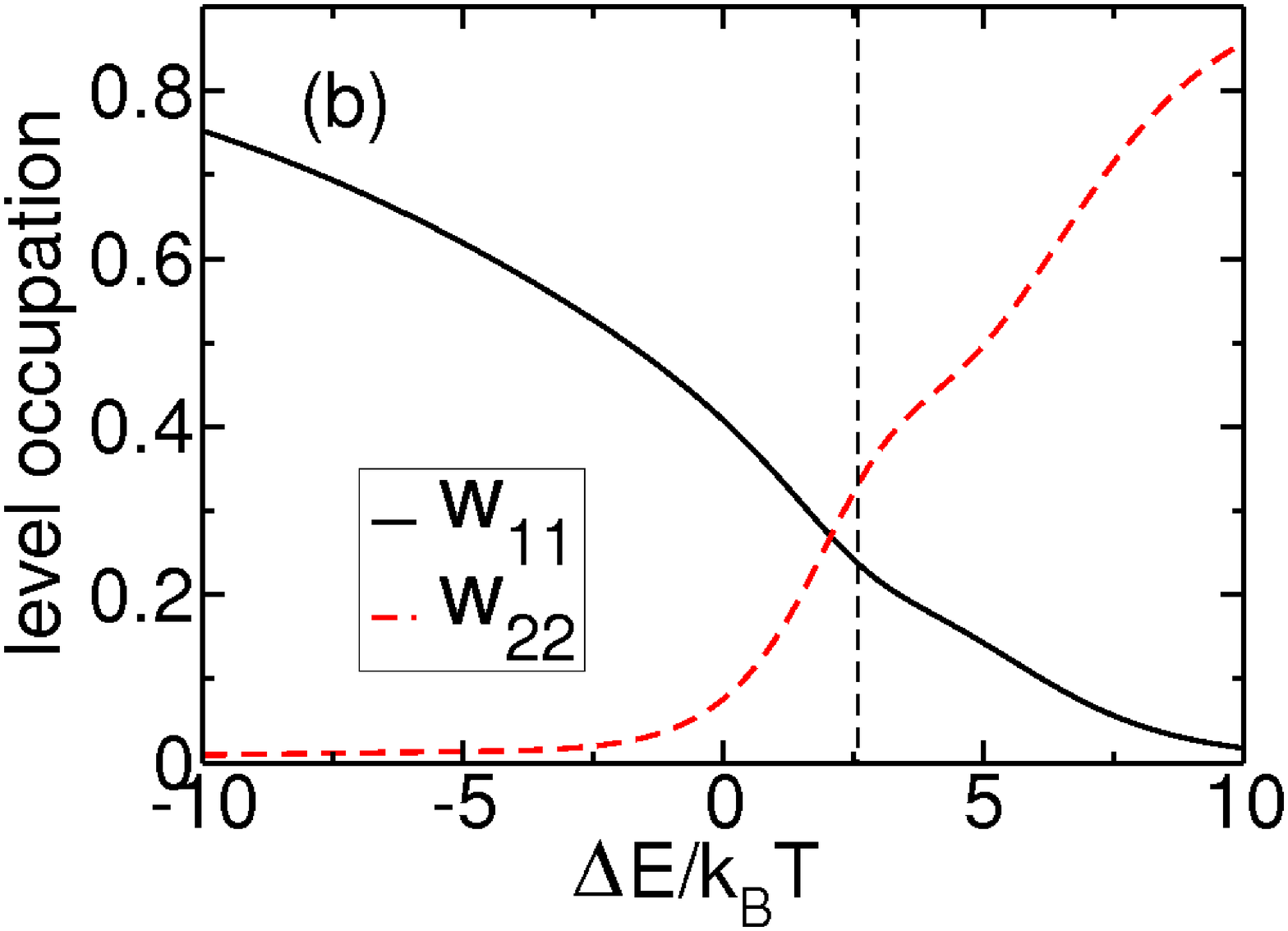}}
\end{minipage}
\begin{minipage}{.45\linewidth}
%\vspace{-0.25cm}
%\vspace{0.1cm}
\resizebox{!}{26mm}{\includegraphics{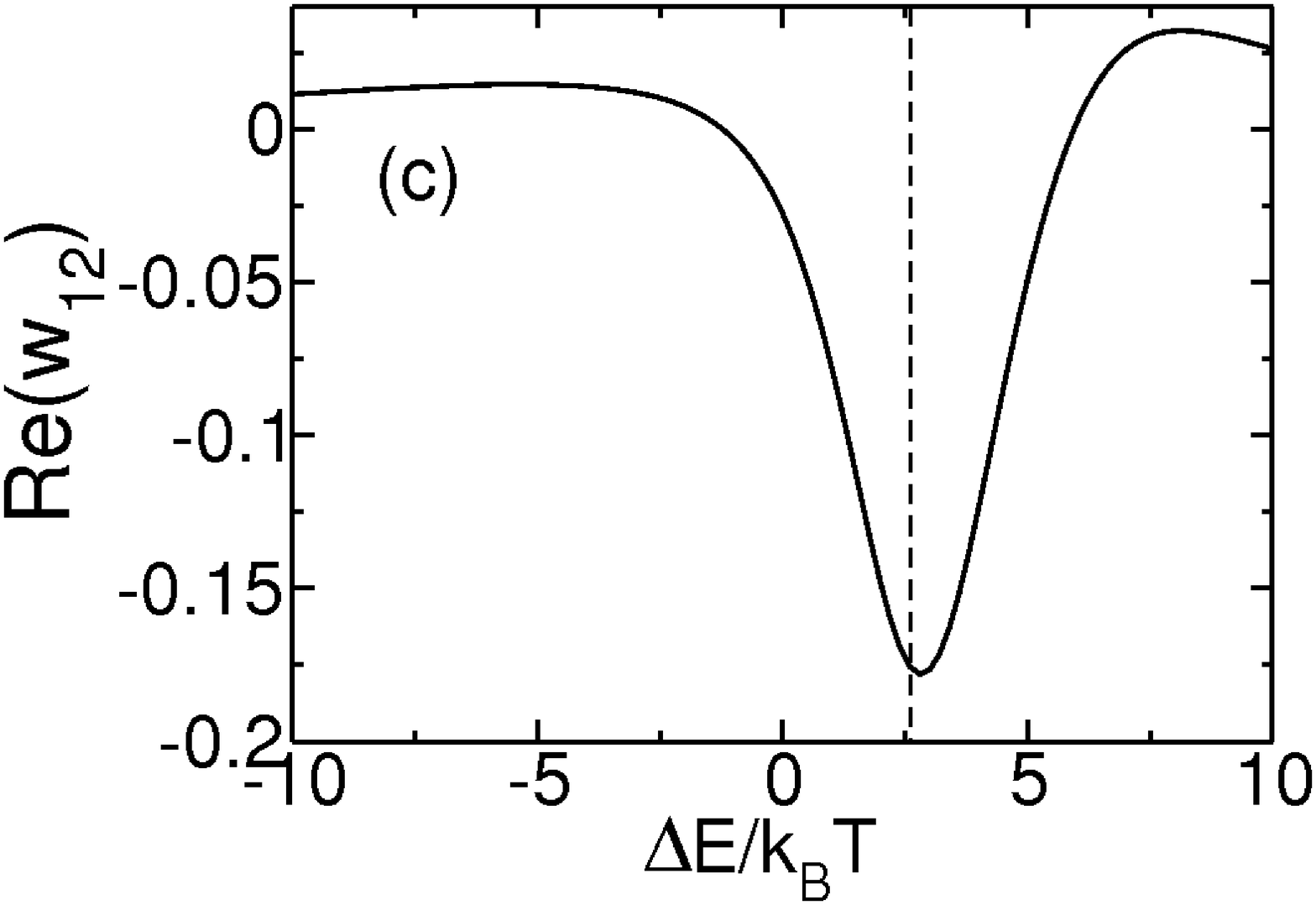}}
\end{minipage}
\end{center}
\caption{(a) The current through the two level quantum dot calculated with 2vN and 1vN formalisms, with and without renormalization of
the energy levels, at $\Gamma_1=4~k_BT$
$U=25 ~k_BT$, $E_g=-U/2$, $a=0.5$ and $V_{\mathrm{bias}}=2.5~k_BT$. As the broadening is not included the first order method 
overestimates the peak height but underestimates the shift away from degeneracy of the conductance dip. (b) and (c) show the 
corresponding occupations and and real part of the coherence calculated with the 2vN formalism. 
The vertical dashed line marks the location of the conductance minimum.}%$k_BT=0.2$
\label{shift}
\end{figure}

With a clear picture of the current suppression at electron-hole symmetry, we now change focus to the regime away from this symmetry 
point. We will investigate the current at $E_g= -U/2$. 
At this point the singly occupied levels are close to chemical potentials of the leads, and sequential tunneling contributes strongly
to the transport. Thus, the 1vN-method gives reasonable results and we can compare first- and second-order von Neumann to investigate
the importance of the broadening in this regime.
Fig.~\ref{shift}(a) shows the current at 
$E_g=-U/2=-12.5~k_BT$  calculated with 2vN and 1vN methods. Since level broadening is not included in the first order method the peaks 
are too high and narrow. 

More interestingly, we see that the current minimum has been shifted away 
from degeneracy. Comparing with the noninteracting results of Fig.~\ref{canyon}, where the dip is found at $\Delta E=0$ at $E_g=\pm U/2$, 
one draws the conclusion that 
the shift is related to the finite charging energy $U$. We will argue that this shift has three origins:

i) A small part of the shift originates from a first order level renormalization and a bias induced anticrossing between the
levels, as can be seen from the slightly different dip location of the two 1vN-curves in Fig.~\ref{shift}a. 
The first order simulations not including these effects also show a shift, 
however slightly smaller. This suggests that there are additional mechanisms producing the shift. The bias induced anticrossing
is further discussed in Sec. VII in connection with the population inversion.

ii) The main shift within the first-order approach can be attributed to the asymmetry in
couplings, which has to be compensated by different spectral weights in order
to allow for full canceling of the individual channels.
As discussed in Ref.~\onlinecite{PhysRevB.71.201308} the spectral density of each state is divided among two peaks, situated at $E$ and $E+U$. 
Increasing the occupation of one level results in a shift in the spectral function of the other level towards the upper peak at $E+U$.
At $E_g=-U/2$ the upper peak is far from the Fermi level of the leads and does not contribute to the transport through the dot.
In order for the stronger/weaker
coupled level (i.e., levels 1/2 for $|a|<1$) to have lower/higher spectral
weight at the Fermi energy, the weaker coupled level  must have a higher
occupation. This is naturally the case, if this level is lower in energy,
thus at $E_g\approx -U/2$ the current minimum requires a positive $\Delta E$, c.f. Eq. (\ref{levels}). 
This effect is reduced for increased bias. For $f_L(E_1)=f_L(E_2)=1$ and $f_R(E_1)=f_R(E_2)=0$, 
complete conductance suppression can be found at degeneracy due to an increased occupation of the weakly coupled level, 
as discussed below.

iii) Comparing the first- and second-order methods it is clear that the inclusion of broadening of the levels increases the shift 
of the conductance dip. It is well known that in a broadened level the low energy part of the spectral density is mostly 
occupied.\cite{BruusBook2004} Due to the Coulomb interaction this results in a significantly higher occupation of the strongly 
coupled level at level degeneracy, see  Fig.~\ref{shift}(b) and Refs.~\onlinecite{SindelPRB2005,MedenPRL2006}. 
To compensate, the minimum is shifted so that the strongly coupled level 
has a higher energy than the weakly coupled. Due to the level broadening one can not expect the two levels to contribute 
equally to the current away from the degeneracy point. As the levels are equally far from the bias window, the broader level will 
have a stronger
contribution if the occupations of the two levels are equal. Thus the minimum can be found for values of $\Delta E$ such that the 
occupation of the weakly coupled level is slightly higher, in agreement with Fig.~\ref{shift}(b). In the regime 
$\mid E_g\mid > U/2$ Coulomb interaction is of minor importance but the dip is still found where the occupation of the weakly coupled 
level is higher, to compensate for the larger broadening of the strongly coupled level. 
This implies the strong shift outside the Coulomb-blockade region, as obtained by NEGF.

Note that the above explanation holds not only in the sequential but also in the co-tunneling regime $-U/2<E_g<U/2$, 
but the argumentation has to be reversed for $E_g>0$, where the current is mainly carried by holes through the doubly occupied states. 
At electron-hole symmetry the singly and doubly occupied levels have the same distance from the bias window, which results in minimal
conductance at level degeneracy.

The canceling of the different current paths through the dot is related to coherence between the levels. For the couplings 
considered here we concluded in Sec.~V that a negative real part of the coherence led to a decreased current. 
It is thus not surprising that Fig.~\ref{shift}(c) shows that the current minimum coincides with a minimum in the 
real part of the coherence.

\begin{figure}[t]
\begin{center}
\resizebox{!}{60mm}{\includegraphics{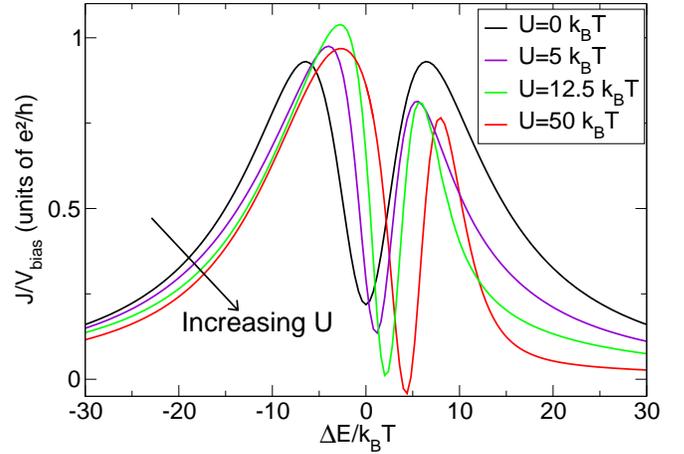}}
\end{center}
\caption{The conductance of the two level quantum dot at  $\Gamma_{1}=5~k_BT$, $a=0.5$, $V_{\mathrm{bias}}=2.5~k_BT$ 
and $E_g=-U/2$. Increasing $U$ leads to a shift of the conductance minimum towards lower energy of the weakly coupled state.}%$k_BT=0.2$
\label{lowbias}
\end{figure}

The relation between the charging energy and the shift of the dip can be better understood by studying the current at
$E_g=-U/2$ for different values of $U$. This is done in Fig.~\ref{lowbias} for $V_{\mathrm{bias}}=2.5~k_BT$. 
Here $U=0$ corresponds to electron-hole symmetry and the conductance suppression is not complete due to the finite bias. As the
charging energy increases, the current will mainly flow through the singly occupied dot states. As predicted above, this results in a 
shift of the conductance dip towards lower energy of the weakly coupled state. Due to the different occupations the current through
the two levels cancel for larger $U$, resulting in a strong conductance suppression. Thus a larger $U$ results in stronger
conductance suppression at the dip, in agreement with Fig.~\ref{lowbias}.

We now focus on the case of higher bias. When both levels are in the bias window and $k_BT\ll V_{\mathrm{bias}}$ the current can be 
calculated using the first-order results. 
Here it is important whether the charging energy is dominating over bias or not. For $U\ll V_{\mathrm{bias}}$, i.e. in the ultra-high-bias 
limit [Fig.~\ref{electronholsym}(c)], both the singly 
and doubly occupied levels are inside the bias window and the current can be calculated using Eq.~(\ref{smallU}), 

\begin{eqnarray}
J=\frac{\Gamma_1}{\hbar}\frac{(1+a^{2})\left[(1-a^{2})^{2}+(E_1-E_2)^{2}/\Gamma_1^{2}\right]}{2\left[(1+a^{2})^{2}+(E_1-E_2)^{2}/\Gamma_1^{2}\right]}.
\label{smallU2}
\end{eqnarray}

\noindent in agreement with Ref.~\onlinecite{SchallerPRB2009}. 
The dip is found at $E_1=E_2$ but the suppression is not complete. (Note that this result corresponds to electron-hole symmetry.)

\begin{figure}[t]
\begin{center}
\resizebox{!}{60mm}{\includegraphics{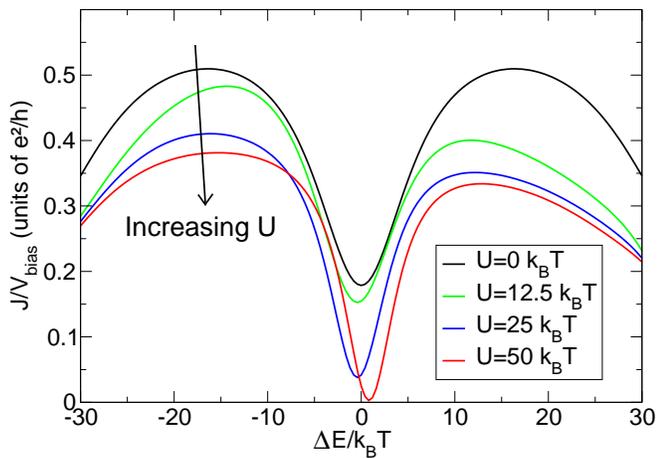}}
\end{center}
\caption{The conductance of the two level quantum dot at  $\Gamma_{1}=5~k_BT$, $a=0.5$, $V_{\mathrm{bias}}=~30 k_BT$ 
and $E_g=-U/2$. As predicted, increasing $U$ leads to reduced conductance at level degeneracy.}%$k_BT=0.2$
\label{Gurvitz}
\end{figure}

For $U\rightarrow \infty$, i.e. the high-bias case [Fig.~\ref{electronholsym}(b)], the doubly occupied state will be empty and the current is given by 
Eq.~(\ref{Gurvitz2}),

\begin{eqnarray}
J=\frac{\Gamma_1}{\hbar}\frac{(1+a^2)(E_1-E_2)^{2}/\Gamma_1^{2}}{(1+a^{2})^{2}+3(E_1-E_2)^{2}/\Gamma_1^{2}},
\label{Gurvitz3}
\end{eqnarray}
 
\noindent again in agreement with Ref.~\onlinecite{SchallerPRB2009}. We note that the current is zero at $E_1=E_2$. This can be explained intuitively:\cite{GurvitzEPL2009,PhysRevB.80.033302}  
At degeneracy we are 
free to work in any linear combination of our original single particle basis. Especially we can choose a basis such that one level, 
$\Psi_{\mathrm{filled}}=\alpha\Psi_1+\beta\Psi_2$, is completely decoupled from the lead with the low chemical potential. 
This level will always be occupied due to the high bias. However, the other level cannot be occupied as $U\rightarrow \infty$. 
As a result neither level can carry any current through the dot, as seen for $U=50~k_BT$ in Fig.~\ref{Gurvitz}. 
It is evident that if level 1 is more strongly coupled than level 2, then $\beta>\alpha$ implying that the weakly
coupled level has a higher occupancy, as discussed above.

At high bias, as in Fig.~\ref{Gurvitz}, the conductance minimum can be found close to level degeneracy in agreement with Eqs.~(\ref{smallU2}) 
and (\ref{Gurvitz3}). The symmetry with respect to mirroring in $\Delta E=0$, for $U=0$ is broken by a 
finite $U$. Comparing Figs.~\ref{lowbias} and \ref{Gurvitz} shows that increasing the bias increases the width of the dip and broadens the peaks.

In this section the current suppression away from electron-hole symmetry was investigated. Since the levels were close to the bias 
window, sequential tunneling was important, and different parity of the two levels was essential for the suppression. At finite bias, 
the suppression was not complete unless the levels had the same coupling strength. However, a finite Coulomb interaction shifted the
dip away from degeneracy and resulted in a stronger suppression. This shift, not previously discussed in the literature, was 
studied in detail. 
Three origins for the shift were found: (i) a renormalization of the energy levels,
(ii) a shift required to give the weakly coupled level a higher occupation which reduces the spectral weight of the singly occupied
state for the other level in the case of Coulomb repulsion, and (iii) a shift 
related to the broadening of the levels. Furthermore, in the high-bias regime, Fig.~\ref{electronholsym}(b), 
there was complete current suppression at the degeneracy point.

\section{CURRENT PEAKS AND POPULATION INVERSION}

\begin{figure}[ht]
\begin{center}
\resizebox{!}{60mm}{\includegraphics{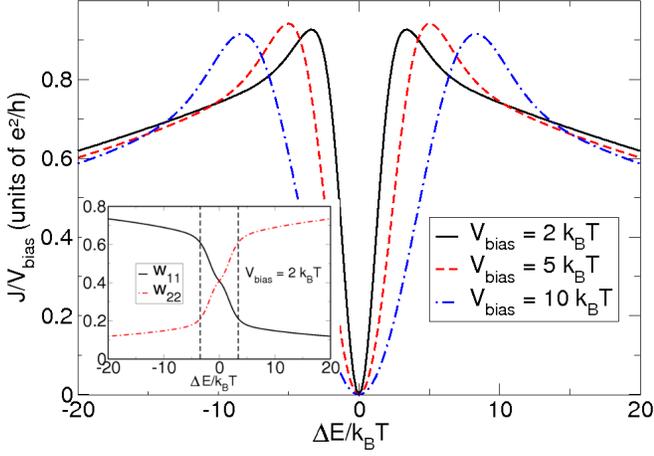}}
\end{center}
\caption{Conductance at $E_g=0$ of the two level quantum dot as a function of the level splitting for different 
values of $V_{\mathrm{bias}}$. The insert shows the occupation of the two single particle levels for $V_{\mathrm{bias}}=2~k_BT$, 
with the vertical dashed lines marking the location of the peaks. The parameters are given by $a=1$, $\Gamma_{1}=20~k_BT$ 
and $U=100~k_BT$.}%$k_BT=0.05$
\label{PeaksBias}
\end{figure}

In this section we change the focus from the dip to the surrounding current peaks. 
In the limit of zero bias and zero temperature and $w_{00}=w_{dd}=0$, it was shown in Ref.~\onlinecite{KashcheyevsPRB2007} that, for the 
couplings of Eq.~(\ref{couplings}), the peaks correspond 
to a difference in occupation of $\vline w_{11}-w_{22}\vline=0.5$. 
To verify if the 2vN method can reproduce these results we must go to the regime $\Gamma\gg V_{\mathrm{bias}},k_BT$, i.e. to the boundary 
of the region of validity of the 2vN-method. To reduce the possible occurrence of negative currents we use couplings with $a=1$, 
where the current is inherently zero at degeneracy. The result of the simulations is shown in Fig.~\ref{PeaksBias}, 
where the black curve 
corresponds to low bias, $V_{\mathrm{bias}}=2~k_BT$. Closer examination reveals that the peaks are located at $\Delta E=\pm3.4~k_BT$ 
where $\vline w_{11}-w_{22}\vline\approx0.4$ in good agreement with the analytical prediction. It is not surprising that the difference 
in occupation is smaller than predicted since there is also a finite probability that the dot is empty or doubly occupied.

\begin{figure}[ht]
\begin{center}
\resizebox{!}{55mm}{\includegraphics{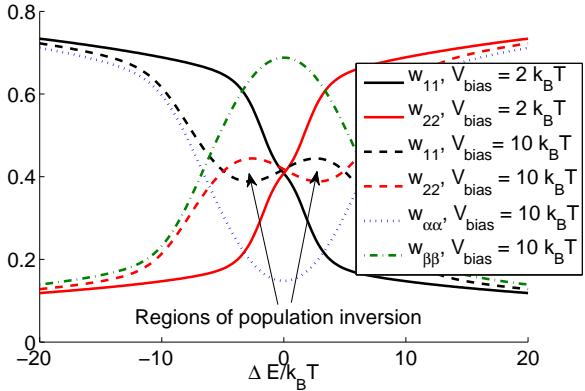}}
\end{center}
\caption{Occupation of the two single particle levels for different values of $V_{\mathrm{bias}}$. Around degeneracy a region of 
population inversion can be found. The parameters are given by $a=1$, $\Gamma_{1}=20~k_BT$ and $U=100~k_BT$. The population inversion in
the diagonal basis ($\left.\vline~\alpha\right\rangle$, $\left.\vline~\beta\right\rangle$) is calculated using $\Omega$ from 
Eq.~(\ref{anticrossing}).}%$k_BT=0.05$
\label{Popinversion}
\end{figure}

A more surprising result is found when the occupations for higher values of $V_{\mathrm{bias}}$ are studied, see Fig.~\ref{Popinversion}. 
This is a regime of validity for the 2vN method due to the increased bias.
It should be pointed out that the regime described here, where the upper state is far below the chemical potentials, 
can not be studied in pure second-order perturbation in $\Gamma$. Such methods do not include mechanisms for emptying the upper 
state.\cite{WeymannPRB2005,BeckerPRB2008}
On each side of the level degeneracy there is a region of population inversion, i.e. the upper level has a higher occupation.
This population inversion has not previously been investigated for our system, however, a similar heating of single electron
transistors due to
inelastic co-tunneling processes was discussed in Ref.~\onlinecite{LaaksoPRL2010}.
The 2vN-simulation for the single-level dot with spin, where a magnetic field splits both spin directions, 
(Anderson model, not shown) reveals no 
population inversion between the two spin levels.
This suggests that the coherence between the two orbital levels is essential for the inversion. The coherence simply means that an 
electron can be in a superposition of the two states, which results from the overlap between the dot levels introduced by the couplings
to the leads. For the level configuration of Fig.~\ref{Popinversion} the dot can as a good approximation be considered to be singly 
occupied, which
means that the Schrieffer-Wolff transformation of Sec. IV can be used to evaluate the overlap. As noted in Sec. IV, in connection
with Eq.~(\ref{lowenergy2}), the contribution
to the overlap from state $E_k$ in the leads is proportional to $c_{kL}^{\dagger}c_{kL}-c_{kR}^{\dagger}c_{kR}\Longrightarrow f_L(E_k)-f_R(E_k)$, which
vanishes in the low bias limit. As a result the coherence and the overlap between the dot state vanishes. However, a finite bias results
in an effective overlap between the two states. Replacing the sums with integrals and using the notation of Eq.~(\ref{levels}) one 
arrives at the following expression for the anticrossing $\Omega$

\begin{eqnarray}
\label{anticrossing}
\Omega=\frac{1}{2} at^{2}\left(\mathrm{ln} \left| \frac{V_{\mathrm{bias}}-U-\Delta E}{V_{\mathrm{bias}}+U+\Delta E} \right|\right. \nonumber \\
\left.+\mathrm{ln} \left| \frac{V_{\mathrm{bias}}-U+\Delta E}{V_{\mathrm{bias}}+U-\Delta E} \right| \right)
\end{eqnarray}

\noindent in the additional term of the dot Hamiltonian given by $\Omega(d_1^{\dagger}d_2+d_2^{\dagger}d_1)$.
We have assumed $k_BT=0$. As long as $\Delta E<U$, which always is the case if Eq.~(\ref{anticrossing}) is valid, 
$\Omega$ is negative for $V_{\mathrm{bias}}>0$. For small bias $V_{\mathrm{bias}}<\Delta E$ the anticrossing vanishes. This is expected as inelastic 
processes where the state of the dot is changed is only possible as long as $V_{\mathrm{bias}}>\Delta E$.
To investigate the effect of this anticrossing we diagonalize 
the dot Hamiltonian. The resulting eigenvalues and eigenvectors are given by

\begin{eqnarray}
E_{\alpha/\beta}=\frac{1}{2}\left(E_1+E_2\mp\sqrt{\Delta E^2+4\Omega^2} \right),
\label{eigenvalues}
\end{eqnarray}

\begin{eqnarray}
\label{eigenvectors2}
\left.\vline~\alpha\right\rangle=x_{\alpha 1}\left.\vline~1\right\rangle+x_{\alpha 2}\left.\vline~2\right\rangle,\\
\left.\vline~\beta\right\rangle=x_{\beta 1}\left.\vline~1\right\rangle+x_{\beta 2}\left.\vline~2\right\rangle,\nonumber
\end{eqnarray}

\noindent where the coefficients are given by

\begin{eqnarray}
\label{eigenvectors}
x_{\alpha 1}=\frac{N_{\alpha}}{2\Omega}\left(\Delta E-\sqrt{\Delta E^2+4\Omega^2}\right),~x_{\alpha 2}=N_{\alpha} \\
x_{\beta 1}=\frac{N_{\beta}}{2\Omega}\left(\Delta E+\sqrt{\Delta E^2+4\Omega^2}\right),~x_{\alpha 2}=N_{\beta} \nonumber
\end{eqnarray}

\noindent respectively, with $N_\alpha$ and $N_\beta$ being normalization constants. Clearly we always have $E_{\alpha}<E_{\beta}$.
The new eigenstates also get new effective couplings

\begin{eqnarray}
t_{L\alpha}=(x_{\alpha 1}t-x_{\alpha 2}at),~t_{R\alpha}=(x_{\alpha 1}t+x_{\alpha 2}at), \nonumber \\
t_{L\beta}=(x_{\beta 1}t-x_{\beta 2}at),~t_{R\beta}=(x_{\beta 1}t+x_{\beta 2}at). 
\label{anticrossing_couplings}
\end{eqnarray}

\begin{figure}[t]
\begin{center}
\resizebox{!}{25mm}{\includegraphics{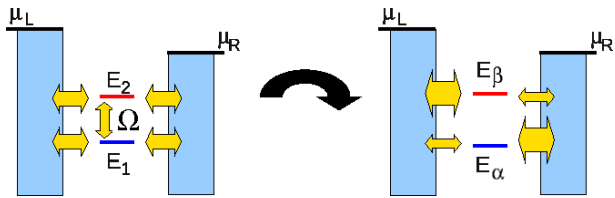}}
\end{center}
\caption{The finite bias induces an anticrossing $\Omega$ between the two original states. Diagonalizing the dot Hamiltonian results
in the upper state being strongly coupled to the lead with high chemical potential, while the lower state is strongly coupled to the
lead with low chemical potential.}
\label{Popinversionfig2}
\end{figure}

\noindent Due to $\Omega<0$ both coefficients $x_{\alpha 1}$ and $x_{\alpha 2}$ of the lower eigenstate 
$\left.\vline~\alpha\right\rangle$ have the same sign. Combined with the difference in 
parity of our two original states this results in a weak coupling of $\left.\vline~\alpha\right\rangle$ to the left lead, i.e. the lead 
with high
chemical potential, and a strong coupling to the right lead. In the same way $\left.\vline~\beta\right\rangle$ couples strongly to the 
left and weakly to the right, see Fig.~\ref{Popinversionfig2}.
The population inversion is real in the sense that it is present in the diagonal basis of the dot 
($\left.\vline~\alpha\right\rangle$, $\left.\vline~\beta\right\rangle$), see Fig.~\ref{Popinversion}. It might at first seem surprising
that the population inversion is so large. For $V_{bias}=10~k_BT$ in Fig.~\ref{Popinversion} we have $\Omega \approx 4~k_BT$ over 
the entire range of $\Delta E$. The detuning $E_\beta-E_\alpha=2\Omega$ at $E_1-E_2=0$, i.e. it is almost the size of $V_{bias}$, and
yet a strong population inversion is observed. The cause for this population inversion is inelastic processes driving a current from
left to right by filling the state $\left.\vline~\beta\right\rangle$ and emptying $\left.\vline~\alpha\right\rangle$. Due to the 
asymmetric couplings of the ($\left.\vline~\alpha\right\rangle$, $\left.\vline~\beta\right\rangle$)-basis such processes are very strong.
The populations of Fig.~\ref{Popinversion} in the ($\left.\vline~\alpha\right\rangle$, $\left.\vline~\beta\right\rangle$)-basis are 
calculated assuming $\Omega$ given by Eq.~(\ref{anticrossing}). Deviations from single occupation of the quantum dot, due to the 
broadening of the dot levels and the finite bias, cause slightly different values of $\Omega$. However, at $\Delta E=0$,
$\left.\vline~\alpha\right\rangle$ and $\left.\vline~\beta\right\rangle$ are the symmetric and antisymmetric combination of
$\left.\vline~1\right\rangle$ and $\left.\vline~2\right\rangle$ respectively, independently of $\Omega$. Therefore, the population
inversion is independent of $\Omega$ at $\Delta E=0$. For higher bias this results in populations close to $w_{\alpha\alpha}=0$ and
$w_{\beta\beta}=1$. In the limit $V_{bias}\rightarrow\infty$, it can not be considered as a population inversion since $\Omega$ vanishes 
in this limit, resulting in $E_{\alpha}=E_{\beta}$.

When $\Delta E<0$, i.e. $E_1<E_2$, $\left.\vline~\alpha\right\rangle$ is mainly composed of the original state 
$\left.\vline~1\right\rangle$, while $\left.\vline~\beta\right\rangle$ is mainly composed of $\left.\vline~2\right\rangle$. 
For $\Delta E>0$ the 
situation is the reverse. Thus, it is always the case that the upper state $\left.\vline~\beta\right\rangle$, that couples strongly 
to the left, is mainly composed of the upper state in the original basis. 
This explains the observed population inversion in our original basis.
A negative bias results in $\Omega>0$, which leads to a strong coupling of
$\left.\vline~\alpha\right\rangle$ to the left. However, for a negative bias the left lead has a low chemical potential. Thus the explanation holds
also for negative bias.

\begin{figure}[t]
\begin{center}
\resizebox{!}{55mm}{\includegraphics{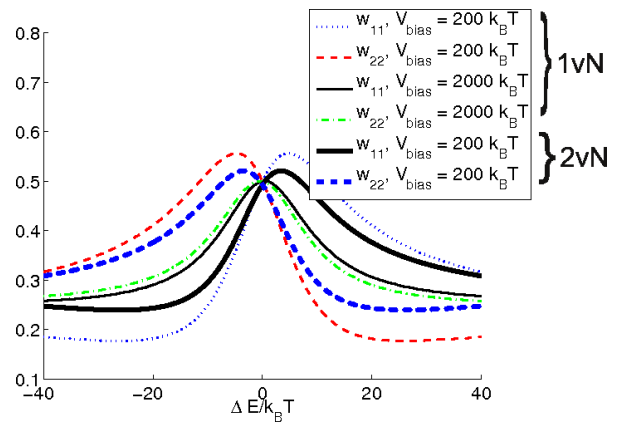}}
\end{center}
\caption{Populations calculated with the 1vN- and 2vN-methods. Couplings and charging energy are given by $\Gamma_{1}=5~k_BT$ and
$U=100~k_BT$, respectively.}
\label{Popinversionfig3}
\end{figure}

The $k=k'$ terms of Eq.~(\ref{lowenergy2}) are present also in first order perturbation in dot-lead couplings. We have seen that the 
real part of the sum, (or integral in the continuous limit), over $k=k'$ gives rise to the population inversion. We thus expect to see 
population inversion also in first order simulations if the real part is included. However, first order perturbation theory can not 
be applied to the co-tunneling regime where Eq.~(\ref{lowenergy2}) is valid. We instead investigate the ultra-high bias regime where
all states are inside the bias window. Eq.~(\ref{vonNeumann}) can be solved including the real part of the sums. The result is shown in 
Fig.~\ref{Popinversionfig3}, where also the result of a 2vN-simulation is shown as comparison. 
One observes that as the bias is increased, which means 
that the real parts of the sums becomes smaller, the occupations approach the infinite bias result of Eq.~(\ref{smallU}).
It is evident that the broadening results in a smaller 
population inversion. We can therefore not completely rule out the possibility that the population inversion is an effect of neglecting
higher order couplings. As the inversion is present also at very high bias, where the contribution from higher order terms is small, we
believe, however, that it is a real effect. Since it is the real part of of the sums of Eq.~(\ref{vonNeumann}) that is responsible for the 
renormalization one would not expect to see any population inversion when these are neglected. Indeed, already at $V_{bias}=200~k_BT$
one finds that the results of such simulations are indistinguishable from Eq.~(\ref{smallU}). These results are therefore not shown
in Fig.~\ref{Popinversionfig3}.

The population inversion is in fact related to the conductance peaks, by increasing the elastic co-tunneling. 
Let us assume that $E_2>E_1$ so that $w_{22}>w_{11}$.
Then in Eq.~(\ref{currentoperator}) the elastic processes with $d_2^{\dagger}d_2$ will dominate. If the levels are placed as in 
Fig.~\ref{electronholsym}(a), 
then $E_k-(E_1+U)$ and $-(E_k-E_2)$ are both positive and there is constructive interference between these tunneling processes. 
Put simply, the population inversion activates elastic processes through the states close to the bias window, i.e. processes that
contribute strongly to the current. Furthermore, the elastic processes have the same phase and interfere constructively.
For large detuning, $\Delta E$, the coherence decreases and as a result the occupation of the upper level will decrease. 
Increased
temperature, which also results in a lower coherence and a related decrease in population inversion, also leads to decreased peak height. 
This scenario describes the formation of the conductance peaks at finite
bias. 

Finally we would like to point out the experimental significance of the results presented in this section. By embedding the quantum dot
in e.g. a microwave cavity the dot can be connected to a photon mode. Due to the population inversion this couping might lead to
stimulated emission of photons, i.e. a maser is formed. Previously it has been suggested that such a population inversion can be reached 
in double quantum dot systems, where the inversion results from an asymmetric couping of the dots to left and right 
lead.\cite{ChildressPRA2004} 
The results of this section suggest that an increased bias will modify the dot Hamiltonian 
of two-level quantum dots so that such couplings are 
achieved, as long as the original dot levels have different parity. 
The difference in parity guarantees a non-zero dipole element between the dot levels. One would expect the dipole element to be larger
in two-level quantum dots than in double quantum dots due to the spatial overlap of the two states. This suggests an increased 
observability of population inversion in two-level quantum dots.

\section{CONCLUDING REMARKS}

A thorough study of the transport through spinless two-level quantum dots has been presented. Regimes not previously studied
in the presence of co-tunneling, such as the regimes of finite bias and temperature, as well as the transport away from electron-hole 
symmetry, was studied. To present a coherent 
picture of the transport through the system, comparison with previously derived results were made.

Close to degeneracy a suppression of the current was found, 
while a slight detuning of the energy levels resulted in current peaks. This resulted in a canyon of current suppression 
cutting through both the sequential and co-tunneling regime. Different mechanism for the suppression were found in the different 
regimes. In the Coulomb blockade region an intuitive explanation to the suppression was presented in the 
low-energy regime, in terms of canceling of different tunneling processes due to equal occupation of the two levels. In the high-bias 
limit with infinite charging energy a complete suppression was found since the electron was trapped in a non-conducting state, preventing
any transport. These two mechanisms for the suppression are independent of the lead-dot couplings. When neither of the above
mechanism were present, different parity of the two dot-levels had to be assumed to achieve current suppression. For such couplings
the canyon was found close to degeneracy in all four cases (high and low bias, strong and 
weak couplings), independent of the level positions, albeit the current did not always entirely drop to zero.  

To quantify the current suppression, simulations were performed using the second-order von Neumann technique. In the regime where the
lead-dot couplings dominate over temperature and applied bias, i.e. co-tunneling processes contribute strongly to the current, 
a strong suppression was observed both in the sequential and the co-tunneling regime, assuming different parity of the two levels. 
For weaker couplings, where co-tunneling could be neglected, only partial
suppression was observed in agreement with analytical results. 

Away from the electron-hole symmetry point a shift of the conductance dip was 
observed. The shift was shown to have three origins: (i) a renormalization of the energy levels,
(ii) a shift required to give the weakly coupled level a higher occupation, and (iii) a shift 
related to the broadening of the levels.

Finally the conductance peaks were investigated. The peaks were found to be intimately related to an increased occupation of the upper
level. The increased population resulted from the bias-induced anticrossing between the two orbital levels.
Sufficiently high bias resulted in an anticrossing large enough to cause population inversion.

%\begin{acknowledgements}
\acknowledgments

We would like to thank Henrik Nilsson, Hongqi Xu and Christian Bergenfeldt for many discussions. Financial support from the 
Swedish Research Counsil (VR) is gratefully acknowledged.
%\end{acknowledgements}

\appendix
\section{CURRENT IN THE FIRST-ORDER von NEUMANN APPROACH}

In this appendix we will derive expressions for the current assuming that the broadening can be neglected. The results are thus of first 
order in $\Gamma$. In the von Neumann formalism the Markov approximation is made within the Schr\"odinger picture, as compared to the
Redfield method where the Markov limit is taken in the interaction picture. This results in slightly different results when the levels
are non-degenerate.\cite{PedersenPRB2007}

Eq.~(\ref{HT}) defines $t_{n\ell}(k)$ as the coupling of a single particle state $n$ to state $k$ in lead $\ell$. The couplings of the 
corresponding many particle states are given by

\begin{eqnarray}
T_{ba\ell}(k)=\sum_{n}t_{n\ell}(k)\left\langle b~\vline d_n^{\dagger}\vline~ a\right\rangle,
\end{eqnarray}

\noindent where $\left.\vline~ a\right\rangle$ and $\left.\vline~ b\right\rangle$ are dot states. 
We use the convention that the many particle state $b$ contains one more dot electron 
than $a$ and so on. Assuming energy independent couplings, $T_{ba\ell}(k)=T_{ba\ell}$, 
the equation of motion for the reduced density matrix is in the first 
order von Neumann method given by:\cite{PedersenPRB2007}

\begin{widetext}
\begin{eqnarray}
\label{vonNeumann}
&&\mathrm{i}\hbar\frac{d}{dt}w_{bb'}=(E_b-E_{b'})w_{bb'}\\
&+&\sum_{a,k\ell}T_{ba\ell}\frac{\sum_{a'}w_{aa'}T_{b'a'\ell}^{*}f_k-\sum_{b''}T_{b''a\ell}^{*}w_{b''b'}(1-f_k)}{E_k-E_{b'}+E_a-\mathrm{i}0^{+}}
-\sum_{a,k\ell}\frac{\sum_{a'}T_{ba'\ell}w_{a'a}f_k-\sum_{b''}w_{bb''}T_{b''a\ell}(1-f_k)}{E_k-E_b+E_a+\mathrm{i}0^{+}}T_{b'a\ell}^{*}\nonumber\\
&+&\sum_{c,k\ell}T_{cb\ell}^{*}\frac{\sum_{b''}T_{cb''\ell}w_{b''b'}f_k-\sum_{c'}w_{cc'}T_{c'b'\ell}(1-f_k)}{E_k-E_c+E_{b'}+\mathrm{i}0^{+}}
-\sum_{c,k\ell}\frac{\sum_{b''}w_{bb''}T_{cb''\ell}^{*}f_k-\sum_{c'}T_{c'b\ell}^{*}w_{c'c}(1-f_k)}{E_k-E_c+E_b-\mathrm{i}0^{+}}T_{cb'\ell},\nonumber
\end{eqnarray}
\end{widetext}

\noindent where $f_k$ is the Fermi distribution.

Going to the continuum limit one replaces the sums over $k$ with integrals. The real parts of the integrals constitute 
a renormalization of the energy levels. However, the results of this Appendix are applied either to electron-hole symmetry or to the 
high/ultra-high-bias case. In these situations the renormalization vanishes, and for the couplings of
Eq.~(\ref{couplings}) the stationary 1vN equations read for the vacuum state $b=0$, $b'=0$:

\begin{widetext}
\begin{eqnarray}
\label{first}
&-&w_{00}\left\lbrace f_L(E_1)+f_{R}(E_1)+a^{2}[f_L(E_2)+f_R(E_2)] \right\rbrace
+w_{11}\left\lbrace 2-f_{L}(E_1)-f_{R}(E_1)\right\rbrace \\
&+&a^{2}w_{22}\left\lbrace 2-f_{L}(E_2)-f_{R}(E_2)\right\rbrace
+a\mathcal{R}\left\lbrace w_{12}\right\rbrace \left\lbrace f_L(E_1)+f_L(E_2)-f_R(E_1)-f_R(E_2)\right\rbrace=0,\nonumber
\end{eqnarray}

for the single particle state $b=1$, $b'=1$:

\begin{eqnarray}
\label{w11}
&&w_{00}\left\lbrace f_{L}(E_1)+f_{R}(E_1) \right\rbrace
-w_{11}\left\lbrace 2-f_{L}(E_1)-f_{R}(E_1)+[f_{L}(E_2+U)+f_{R}(E_2+U)]a^{2} \right\rbrace\\
&+&a^{2}w_{dd}\left\lbrace 2-f_{L}(E_2+U)-f_{R}(E_2+U) \right\rbrace
-a\mathcal{R}\left\lbrace w_{12} \right\rbrace \left\lbrace f_{L}(E_1)+f_{L}(E_2+U) -f_{R}(E_1)-f_{R}(E_2+U)\right\rbrace=0,\nonumber
\end{eqnarray}

for the other single particle state $b=2$, $b'=2$:

\begin{eqnarray}
\label{last}
&&a^{2}w_{00}\left\lbrace f_{L}(E_2)+f_{R}(E_2) \right\rbrace
-w_{22}\left\lbrace [2-f_{L}(E_2)-f_{R}(E_2)]a^{2}+f_{L}(E_1+U)+f_{R}(E_1+U) \right\rbrace \\
&+&w_{dd}\left\lbrace 2-f_{L}(E_1+U)-f_{R}(E_1+U) \right\rbrace
-a\mathcal{R}\left\lbrace w_{12} \right\rbrace \left\lbrace f_{L}(E_2)+f_{L}(E_1+U) -f_{R}(E_2)-f_{R}(E_1+U)\right\rbrace=0,\nonumber
\end{eqnarray}

and for the coherence $b=1$, $b'=2$, ($b=2$, $b'=1$ gives the same information since $w_{21}=w_{12}^{*}$):

\begin{eqnarray}
\label{last2}
&-&\mathrm{i}a\frac{w_{00}}{2}\left\lbrace f_{L}(E_1)+f_L(E_2)-f_{R}(E_1)-f_{R}(E_2)\right\rbrace
-\mathrm{i}a\frac{w_{dd}}{2}\left\lbrace f_{L}(E_1+U)+f_{L}(E_2+U)-f_{R}(E_1+U)-f_{R}(E_2+U)\right\rbrace \nonumber\\
&-&\mathrm{i}a\frac{w_{11}}{2}\left\lbrace f_{L}(E_1)-f_{R}(E_1)+f_{L}(E_2+U)-f_{R}(E_2+U)\right\rbrace
-\mathrm{i}a\frac{w_{22}}{2}\left\lbrace f_{L}(E_2)-f_{R}(E_2)+f_{L}(E_1+U)-f_{R}(E_1+U)\right\rbrace \nonumber\\
&-&\mathrm{i}\frac{w_{12}}{2}[2-f_L(E_2)-f_R(E_2)+f_L(E_2+U)+f_R(E_2+U) \\ 
&+&a^{2}(2-f_L(E_1)-f_R(E_1)+f_L(E_1+U)+f_R(E_1+U))]+\frac{E_1-E_2}{\Gamma_1}w_{12}=0. \nonumber
\end{eqnarray}

\end{widetext}

\noindent The equation for $b=d$, $b'=d$ (doubly occupied state) does not provide any new information. One must instead use the normalization

\begin{eqnarray}
w_{00}+w_{11}+w_{22}+w_{dd}=1.
\label{normalization}
\end{eqnarray}

\noindent At $E_1=E_2=-U/2$, where $w_{00}=w_{dd}$, the equations of motion are symmetric in the two levels. This results in an equal population
of the two states. Furthermore the $(E_1-E_2)$-term of Eq.~(\ref{last2}) disappears, implying that the coherence is real
as remarked in Sec.~V.

Once the system of equations has been solved, the current can be calculated with\cite{PedersenPRB2007}

\begin{widetext}
\begin{eqnarray}
J=\frac{\Gamma_1}{\hbar}\left\lbrace w_{00}[f_{L}(E_1)+f_L(E_2)a^{2}]-w_{11}[1-f_{L}(E_1)]+w_{11}f_L(E_2+U)a^{2}-w_{22}[1-f_{L}(E_2)]a^{2}+w_{22}f_L(E_1+U)\right.\nonumber \\
\left.-w_{dd}[1-f_L(E_1+U)+[1-f_L(E_2+U)]a^{2}] +a\mathcal{R}\left\lbrace w_{12}\right\rbrace[2-f_{L}(E_1)-f_L(E_2)+f_{L}(E_1+U)+f_{L}(E_2+U)]\right\rbrace .~~
\label{current2}
\end{eqnarray}
\end{widetext}

We first consider the ultra-high-bias limit, i.e. $f_L(E_1)=f_L(E_2)=f_L(E_1+U)=f_L(E_2+U)=1$ and 
$f_R(E_1)=f_R(E_2)=f_R(E_1+U)=f_R(E_2+U)=0$. Inserting these occupations in Eqs.~(\ref{first})-(\ref{last2}) one obtains

\begin{eqnarray}
\label{A9}
-w_{00}(1+a^{2})+w_{11}+a^{2}w_{22}
+2a\mathcal{R}\left\lbrace w_{12}\right\rbrace&=&0,~~~~~\\
w_{00}-w_{11}(1+a^{2})+a^{2}w_{dd}
-2a\mathcal{R}\left\lbrace w_{12}\right\rbrace&=&0,~~~~~\\
a^{2}w_{00}-w_{22}(1+a^{2})+w_{dd}
-2a\mathcal{R}\left\lbrace w_{12}\right\rbrace&=&0,~~~~~\\
\label{w12}
-\mathrm{i}aw_{00}-\mathrm{i}aw_{11}-\mathrm{i}aw_{22}-\mathrm{i}aw_{dd}&& \nonumber\\
-\mathrm{i}(1+a^{2})w_{12}+\frac{E_1-E_2}{\Gamma_1}w_{12}&=&0.
\end{eqnarray}

\noindent The imaginary part of $w_{12}$ only occurs in the last equation and can easily be 
expressed in terms of the real part of $w_{12}$.

\begin{eqnarray}
\mathcal{I}\left\lbrace w_{12}\right\rbrace =-\frac{(E_1-E_2)\mathcal{R}\left\lbrace w_{12}\right\rbrace}{\Gamma_1(1+a^{2})}.
\end{eqnarray}

\noindent Eq.~(\ref{w12}) can then be rewritten as

\begin{eqnarray}
&&aw_{00}+aw_{11}+aw_{22}+aw_{dd} ~~~~~~~\nonumber \\
&+&\left[ 1+a^{2}+\frac{(E_1-E_2)^{2}}{\Gamma_1(1+a^{2})}\right]\mathcal{R}\left\lbrace w_{12}\right\rbrace=0.
\end{eqnarray}

\noindent Solving for the occupations one finds 

\begin{eqnarray}
w_{11}=w_{22}=\frac{1}{4}+\frac{a^{2}}{(1+a^{2})^{2}+(E_1-E_2)^{2}/\Gamma_1^{2}},
\label{occupation}
\end{eqnarray}

\noindent as noted in Sec.~VII. Using Eq.~(\ref{current2}) the current can be calculated as

\begin{eqnarray}
J=\frac{\Gamma_1}{\hbar}\left[w_{00}(1+a^{2})+w_{11}a^{2} \right. \nonumber \\
\left.+w_{22}+2a\mathcal{R}\left\lbrace w_{12}\right\rbrace\right].
\end{eqnarray}

\noindent Solving Eqs.~(\ref{A9})-(\ref{w12}) results in

\begin{eqnarray}
J=\frac{\Gamma_1}{\hbar}\frac{(1+a^{2})\left[(1-a^{2})^{2}+(E_1-E_2)^{2}/\Gamma_1^{2}\right]}{2\left[(1+a^{2})^{2}+(E_1-E_2)^{2}/\Gamma_1^{2}\right]}.
\label{smallU}
\end{eqnarray}
\vspace{0.1cm}

\noindent It is easy to show that the conductance minimum is found at $E_1=E_2$ where, for a general value of the bias, 
the current is given by Eq.~(\ref{easycurrent}).

We next turn to the high-bias limit $U=\infty$, i.e. $f_L(E_1)=f_L(E_2)=1$ and $f_R(E_1)=f_R(E_2)=f_R(E_1+U)=f_R(E_2+U)=f_L(E_1+U)=f_L(E_2+U)=0$. 
Here it trivially follows from the equation for $b=b'=d$ that $w_{dd}$ is zero. Furthermore Eq.~(\ref{first}) is redundant. 
The remaining equations for the occupations are given by

\begin{eqnarray}
w_{00}-w_{11}-a\mathcal{R}\left\lbrace w_{12} \right\rbrace&=&0,\\
a^{2}w_{00}-a^{2}w_{22}-a\mathcal{R}\left\lbrace w_{12} \right\rbrace&=&0,\\
-\mathrm{i}aw_{00}-\mathrm{i}a\frac{w_{11}}{2}-\mathrm{i}a\frac{w_{22}}{2}&&\nonumber\\
-\mathrm{i}\frac{w_{12}}{2}(1+a^{2})+\frac{E_1-E_2}{\Gamma_1}w_{12}&=&0.\label{w12b}
\end{eqnarray}

\noindent From Eq.~(\ref{w12b}) $\mathcal{I}\left\lbrace w_{12}\right\rbrace$ can be determined

\begin{eqnarray}
\mathcal{I}\left\lbrace w_{12}\right\rbrace =-\frac{2(E_1-E_2)\mathcal{R}\left\lbrace w_{12}\right\rbrace}{\Gamma_1(1+a^{2})}.
\end{eqnarray}

\noindent Thus in the case $U=\infty$ the current is given by

\begin{eqnarray}
J&=&\frac{\Gamma_1}{\hbar}w_{00}(1+a^{2})~~~~\nonumber \\
&=&\frac{\Gamma_1}{\hbar}\frac{(1+a^2)(E_1-E_2)^{2}/\Gamma_1^{2}}{(1+a^{2})^{2}+3(E_1-E_2)^{2}/\Gamma_1^{2}},
\label{Gurvitz2}
\end{eqnarray}

\noindent Once again the conductance minimum is at degeneracy but for $U=\infty$ there is complete conductance suppression independently 
of the relative coupling strength $a$, in agreement with Refs.~\onlinecite{GurvitzEPL2009,PhysRevB.80.033302}.

\section{Renormalization of the dot-levels from the Schrieffer-Wolff transformation}

In this appendix we give the derivation of Eq.~(\ref{renormalization}).
The renormalization, given by the $k=k'$-terms, reads

\begin{widetext}
\begin{eqnarray}
\hat{H}_{\mathrm{renorm}}=&2&\left(\sum_{k}\frac{-t^{2}}{E_k-E_1}(1-\frac{1}{2}(c_{kL}^{\dagger}c_{k_L}+c_{kR}^{\dagger}c_{k_R}))\right.
\left.+\sum_{k}\frac{a^{2}t^{2}}{E_k-(E_2+U)}\frac{1}{2}(c_{kL}^{\dagger}c_{k_L}+c_{kR}^{\dagger}c_{k_R})\right)d_1^{\dagger}d_1 \nonumber \\
+&2&\left(\sum_{k}\frac{-a^{2}t^{2}}{E_k-E_2}(1-\frac{1}{2}(c_{kL}^{\dagger}c_{k_L}+c_{kR}^{\dagger}c_{k_R}))\right.
\left.+\sum_{k}\frac{t^{2}}{E_k-(E_1+U)}\frac{1}{2}(c_{kL}^{\dagger}c_{k_L}+c_{kR}^{\dagger}c_{k_R})\right)d_2^{\dagger}d_2, \nonumber
\end{eqnarray}
\end{widetext}

\noindent where the factor of $2$ accounts for the Hermitean conjugate in Eq.~(\ref{lowenergy2}). 
One sees that the sums are actually logarithmically divergent, which of course seems like a problem at first sight. However, a common
renormalization of the two levels is just a shift of the zero point in the Hamiltonian, that can be neglected. The physically interesting
quantity is the difference in renormalization of the two levels. We first assume that level 2 is fixed, transferring the whole
renormalization to level 1. Replacing the operators with their expectation values $c_{kl}^{\dagger}c_{k_l}\approx\langle c_{kL}^{\dagger}c_{k_L} \rangle=f_l(E_k)=f(E_k)$, 
in the low-bias limit, and using $\Gamma_1(E)=2\pi\sum_k t^{2}\delta(E_k-E)$ and a constant density of states, 
the renormalization is given by

\begin{widetext}
\begin{eqnarray}
\hat{H}_{\mathrm{renorm}}&=&\frac{\Gamma_1}{\pi}\left(-\mathcal{P}\left\lbrace \int_{-\infty}^{\infty}dE_k \frac{1}{E_k-E_1}(1-f(E_k)) \right\rbrace \right.
-\mathcal{P}\left\lbrace \int_{-\infty}^{\infty}dE_k \frac{1}{E_k-(E_1+U)}f(E_k) \right\rbrace \\
&+&a^{2}\mathcal{P}\left\lbrace \int_{-\infty}^{\infty}dE_k \frac{1}{E_k-E_2}(1-f(E_k)) \right\rbrace 
\left.+a^{2}\mathcal{P}\left\lbrace \int_{-\infty}^{\infty}dE_k \frac{1}{E_k-(E_2+U)}f(E_k) \right\rbrace \right)d_1^{\dagger}d_1,\nonumber
\end{eqnarray}
\end{widetext}

\noindent with $\mathcal{P}$ denoting the principal value of the integral. 
Inserting $f(E_k)=1$ for $E_k<0$ and $f(E_k)=0$ for $E_k>0$, corresponding to the low temperature limit, and adding 
the constant term $a^{2}\mathrm{ln}\lvert E_2/(E_2+U)\rvert$ we obtain Eq.~(\ref{renormalization})

\begin{eqnarray}
\hat{H}_{\mathrm{renorm}}=\frac{\Gamma_1}{\pi}\left(\mathrm{ln}\left| \frac{E_1}{E_1+U} \right|d_1^{\dagger}d_1\right. \nonumber \\
\left.+a^{2}\mathrm{ln}\left| \frac{E_2}{E_2+U} \right|d_2^{\dagger}d_2\right),
\end{eqnarray}

\noindent where we have used $d_1^{\dagger}d_1+d_2^{\dagger}d_2=1$, which is applicable due to the single occupation of the dot. 

This expression can also be derived from first-order Redfield which is similar to first order von Neumann,
except that the Markov limit is taken in the interaction picture, while the Schr\"odinger picture is used in 
the von Neumann method. Redfield generates an equation of motion like Eq.~(\ref{vonNeumann}), but with different energy denominators in the 
equations for the off-diagonal terms of the reduced density matrix.

%\bibliography{references_submit}

\end{document}